%% file: main.tex
\documentclass[sigconf, nonacm]{acmart}

\usepackage{pvldb}

\renewcommand\vldbdoi{XX.XX/XXX.XX}
\renewcommand\vldbpages{XXX-XXX}
\renewcommand\vldbavailabilityurl{https://github.com/JesseComer/whiel}

\input{admin/preamble}

\allowdisplaybreaks

\AtBeginDocument{%
  }

\begin{document}

\title{
  \textsc{Whiel}: A Neurosymbolic Platform for Certified
  \texorpdfstring{\\}{ }
  Relational Program Verification%
}

\author{Jesse Comer}
\orcid{0009-0006-9734-3457}
\authornote{The first three student authors are listed alphabetically by last name with their individual contributions detailed in the appendix.}
\affiliation{%
  \institution{University of Pennsylvania}
  \city{Philadelphia}
  \state{Pennsylvania}
  \country{USA}}
\email{jacomer@engineering.upenn.edu}

\author{Fangzhu Shen}
\orcid{0009-0007-3820-5991}
\authornotemark[1]
\affiliation{%
 \institution{Duke University}
 \city{Durham}
 \state{North Carolina}
 \country{USA}}
\email{fangzhu.shen@duke.edu}

\author{Shuo Zhang}
\orcid{0009-0003-6924-6658}
\authornotemark[1]
\affiliation{%
 \institution{Duke University}
 \city{Durham}
 \state{North Carolina}
 \country{USA}}
\email{shuo.zhang2@duke.edu}

\author{Mayur Naik}
\orcid{0000-0003-1348-8618}
\affiliation{%
  \institution{University of Pennsylvania}
  \city{Philadelphia}
  \state{Pennsylvania}
  \country{USA}}
\email{mhnaik@engineering.upenn.edu}

\author{Sudeepa Roy}
\orcid{0009-0002-8300-7891}
\affiliation{%
 \institution{Duke University}
 \city{Durham}
 \state{North Carolina}
 \country{USA}}
\email{sudeepa@cs.duke.edu}

\author{Val Tannen}
\orcid{0009-0008-6847-7274}
\affiliation{%
  \institution{University of Pennsylvania}
  \city{Philadelphia}
  \state{Pennsylvania}
  \country{USA}}
\email{val@engineering.upenn.edu}

\renewcommand{\shortauthors}{Jesse Comer, Fangzhu Shen, Shuo Zhang, Mayur Naik, Sudeepa Roy, \& Val Tannen}

\begin{abstract}
    \input{sections/s0-abstract.tex}
\end{abstract}

\maketitle

\vldbtopmatter

\input{sections/s1-intro}
\input{sections/s2-hoare-validity}
\input{sections/s3-applications}
\input{sections/s4-verification}
\input{sections/s5-system}
\input{sections/s6-experiments}
\input{sections/s7-conclusion}

\bibliographystyle{ACM-Reference-Format}
\bibliography{admin/bib}

\newpage
\appendix
\input{sections/appendix}
\input{sections/s8-contributions}
\end{document}

%% file: admin/preamble.tex
\usepackage{algorithm}
\usepackage[noend]{algpseudocode}
\usepackage{amsmath}
\usepackage{amsthm}
\usepackage{bm}
\usepackage{amsfonts}
\usepackage[bold,full]{complexity}
\usepackage{graphicx}
\usepackage{hyperref}
\usepackage{mathtools}
\usepackage{multicol}
\usepackage{tikz}
    \usetikzlibrary{arrows,shapes,positioning, automata}
    \usetikzlibrary{arrows,automata,positioning}
    \usetikzlibrary{decorations.markings}
    \usetikzlibrary{arrows.meta}
    \usetikzlibrary{backgrounds,fit,calc}
\usepackage[utf8]{inputenc}
\usepackage{xcolor}
\usepackage{colortbl}
\usepackage[normalem]{ulem}
\usepackage{xspace}
\usepackage[shortlabels]{enumitem}
\usepackage{subcaption}
\usepackage{stfloats}

\newcommand{\WhielSkills}{\Whiel{}+Skills\xspace}
\newcommand{\WhielTools}{\Whiel{}+Tools\xspace}
\newcommand{\WhielToolsSkills}{\Whiel{}+Tools+Skills\xspace}

\newcommand{\Ascr}{\mathcal{A}}
\newcommand{\Kscr}{\mathcal{K}}

\newcommand{\Whiel}{\textsc{Whiel}\xspace}
\newcommand{\Lean}{\textsc{Lean}\xspace}
\newcommand{\Rocq}{\textsc{Rocq}\xspace}
\newcommand{\Coq}{\textsc{Coq}\xspace}
\newcommand{\Vampire}{\textsc{Vampire}\xspace}
\newcommand{\VampLean}{\textsc{VampLean}\xspace}
\newcommand{\SQL}{\textsc{SQL}\xspace}
\newcommand{\Datalog}{\textsc{Datalog}\xspace}
\newcommand{\CoreGQL}{\textsc{CoreGQL}\xspace}
\newcommand{\LOOPY}{\textsc{LOOPY}\xspace}
\newcommand{\SQLSolver}{\textsc{SQLSolver}\xspace}
\newcommand{\QED}{\textsc{QED}\xspace}
\newcommand{\VeriEQL}{\textsc{VeriEQL}\xspace}
\newcommand{\Cosette}{\textsc{Cosette}\xspace}
\newcommand{\GraphQE}{\textsc{GraphQE}\xspace}
\newcommand{\Graphiti}{\textsc{Graphiti}\xspace}
\newcommand{\Cypher}{\textsc{Cypher}\xspace}
\newcommand{\Logos}{\textsc{Logos}\xspace}

\newcommand{\directlean}{\textsc{DirectLean}\xspace}
\newcommand{\whieloneround}{\textsc{Whiel-OneRound}\xspace}

\newcommand{\LH}{\textsc{LeveledHoudini}\xspace}
\newcommand{\Houdini}{\textsc{Houdini}\xspace}
\newcommand{\Extend}{\textsc{Extend}\xspace}
\newcommand{\valid}{\texttt{valid}}
\newcommand{\Rold}{R_{old}}
\newcommand{\AgentHoudini}{\textsc{AgentHoudini}\xspace}
\newcommand{\PushTask}{\textsc{PushTask}\xspace}
\newcommand{\Propose}{\textsc{Propose}\xspace}
\newcommand{\Validate}{\textsc{Validate}\xspace}
\newcommand{\Pending}{\mathsf{Pending}}
\newcommand{\Submission}{\mathsf{Submission}}
\newcommand{\LeanCheck}{\textsc{LeanCheck}\xspace}

\newcommand{\dlog}{\;\leftarrow\;}
\newcommand{\IDB}{\operatorname{IDB}}
\newcommand{\EDB}{\operatorname{EDB}}

\newcommand{\Hoare}[3]{\{#1\}\,#2\,\{#3\}}
\newcommand{\init}{\textbf{(init)}\xspace}
\newcommand{\maint}{\textbf{(maint)}\xspace}
\newcommand{\term}{\textbf{(term)}\xspace}
\newcommand{\step}{\textbf{(step)}\xspace}
\renewcommand{\wp}{\textbf{wlp}\xspace}

\newcommand{\FOL}{\textrm{\upshape FOL}\xspace}

\newcommand{\RA}{\textrm{\upshape RA}\xspace}
\newcommand{\select}{\sigma}
\newcommand{\project}{\pi}
\newcommand{\true}{\mathsf{true}}
\newcommand{\false}{\mathsf{false}}

\newcommand{\seq}{~;~}
\newcommand{\sk}{\textbf{skip}\xspace}
\newcommand{\while}{\textbf{while}\xspace}
\newcommand{\whileargs}[2]{\ensuremath{\mathbf{while}\ #1\ \mathbf{do}\ #2\ \mathbf{end}}\xspace}
\newcommand{\ite}{\textbf{if-then-else}\xspace}
\newcommand{\iteargs}[3]{\ensuremath{\mathbf{if}\ #1\ \mathbf{then}\ #2\ \mathbf{else}\ #3}\xspace}

\newcommand{\tcwhiel}{\textbf{TC$_\Whiel$}\xspace}
\newcommand{\Itc}{\ensuremath{I_{\textbf{TC}}}}
\newcommand{\Atcl}{\ensuremath{\Ascr_{\textbf{TC}}}}

\newcommand{\psdatalog}{\textbf{PS$_\Datalog$}\xspace}
\newcommand{\psnaive}{\textbf{PS$_{\textsc{Naive}}$}\xspace}
\newcommand{\psseminaive}{\textbf{PS$_{\textsc{SemiNaive}}$}\xspace}
\newcommand{\rgwhiel}{\textbf{RG$_\Whiel$}\xspace}
\newcommand{\LLMsolve}{LLM$_{\text{Solve}}$}
\newcommand{\LLMwhiel}{LLM$_{\text{\Whiel}}$}
\newcommand{\ivmwhiel}{\textbf{IVM$_\Whiel$}}

%% file: sections/s0-abstract.tex
We present the \Whiel (a contraction of Whi(le-R)el) system for \Lean-certified reasoning about imperative relational programs.
The system can encode containment and equivalence problems for \Datalog, its extensions with negation, and the core fragment of graph query languages.
It can also represent optimized or incremental evaluation algorithms that are not expressible in the original query language, as well as first-order specifications of integrity constraints, schema mappings, and views arising in data integration, exchange, and sharing.
\Whiel's program states are database instances, and its imperative structure enables compositional Hoare-style reasoning with first-order relational specifications, using prophecy variables and leveled invariants to support inductive arguments.
\Whiel supports neurosymbolic discovery of loop invariants whose verification conditions can be checked using the \Vampire first-order theorem prover.
Counterexamples are fed back into the neurosymbolic pipeline, while proposed solutions are translated into Lean proof certificates.
To validate the \Whiel system, we evaluate its performance on a curated benchmark suite.
In summary, we envision the \Whiel system as a tool for tackling mathematically demanding, typically undecidable problems involving complex query and constraint languages, with a high degree of trust in its output.

%% file: sections/s1-intro.tex
\section{Introduction}
\label{sec:intro}

\input{figures/system-overview}

The rapidly growing capabilities of LLMs introduce numerous opportunities for applications in the management of data. A database developer might use an LLM to generate an optimized evaluation strategy for a recursive query, an algorithm for incrementally updating views, or a query rewritten using views. A correct solution to such requests can improve query evaluation, reduce computation as a database is updated, or enable applications to work across schemas in data integration, exchange, and sharing. Fernandez et al.~\cite{fernandez2023how} predict that LLMs will revolutionize how we approach such hard, often undecidable, database problems.

But how can a developer be sure that a solution proposed by an LLM to such problems is correct? An incorrect query rewrite or faulty view maintenance algorithm can lead to wrong or missing answers, which can have serious consequences for users, who depend on the accuracy of query results. One approach is to manually review the LLM-proposed solution; another is to test the query or view maintenance algorithm on sample databases. However, while such approaches might expose mistakes, they do not establish correctness of a rewritten query across all databases. Furthermore, an argument showing that even a short recursive program computes the right results can require a substantial amount of mathematical sophistication. Evidence~\cite{demirer2026writing} suggests that coding agents leave a bottleneck at the point of human review; that is, the burden has shifted from \emph{writing code} to \emph{confirming that the code is correct}. In the case of checking LLM-generated solutions to hard database problems, the time and expertise required for such a review can be impractical. This leaves developers with the choice to either expend the requisite effort, or to trust a solution which has not been independently confirmed.

However, recent successes in AI-assisted mathematics~\cite{openai2026navierstokes} point toward another approach: asking the LLM to \emph{prove} correctness of its solution in an interactive theorem prover such as \Lean~\cite{moura2021lean}. \Lean provides a small, trusted \emph{kernel}; given a carefully formalized theorem statement and a \emph{proof term}, the kernel checks whether or not the proof is valid. Given such a formalization, a human reviewer needs only to inspect the mathematical definitions and theorem statements. However, producing these definitions on each invocation of an LLM to solve a hard database problem means checking the formalization each time, leaving a still substantial burden on the reviewer. Furthermore, such requests can require highly capable models, resulting in significant financial cost in order to produce a solution. One idea to reduce this time and financial cost is to produce reusable definitions one time; however, it still leaves success or failure up to the LLM's ability to construct proofs across a wide range of problem-specific formalizations.

We believe that a simpler approach is to target a single general and hard problem into which all of the others -- query equivalence, containment, rewriting, incremental updating, and more -- can be encoded. As this single general problem we choose \emph{verification of imperative relational programs}. Program verification is a well-studied area with a developed mathematical theory and powerful tools, such as SAT/SMT solvers and automated and interactive theorem provers, capable of coping practically with the intractable and undecidable nature of the problem. Using such tools, instead of a variety of insights for proving the correctness of proposed solutions to a variety of problems, we only need insights for 
``guessing'' loop invariants. We use LLMs for this and then our platform can ``check'' these guesses and certify the correctness proofs.

\subsection*{Our Contributions}

We introduce the \Whiel system, a neurosymbolic platform for certified verification of \emph{relational programs}. In \Whiel's relational programming language, \emph{program states} are database instances, updated by assigning the results of relational algebra (\RA) expressions to relation-valued variables. The system is designed to solve the \emph{\Whiel Hoare validity problem} which, informally, asks whether or not a relational program satisfies a given first-order specification of its correctness (overall formalized as a \emph{Hoare triple}; see Equation~\ref{eq:hoare-triple}). The \Whiel Hoare validity problem serves as the common target proposed above: it is highly general, capable (see Section~\ref{sec:applications}) of simulating not only recursive queries written in \Datalog~\cite{ceri1989datalog}, in its extensions with negation~\cite{abiteboul1995foundations}, or in the \emph{cores} of graph query languages~\cite{gheerbrant2025gql}, but also their evaluation strategies. It can also represent \Datalog containment and equivalence with respect to constraints, optimized or incremental algorithms, correctness of view-based query rewrites, schema mappings, and problems arising in data integration, exchange, and sharing. Through encodings, \Whiel aligns with the vision of Castro Fernandez et al.~\cite{fernandez2023how}, who predicted that LLMs might overcome the ``ceiling'' for existing automation of these hard problems. \Whiel also addresses one of the key challenges of this vision: ensuring that the LLM's proposed solutions are correct.

The primary challenge in solving the \Whiel Hoare validity problem is the synthesis of \emph{loop invariants}: formal assertions used to prove correctness properties of program loops. By pre-processing an arbitrary program written in \Whiel's relational language to \emph{single-loop form}, \Whiel reduces this issue to the problem of synthesizing a single invariant for the pre-processed program. It then implements an LLM-powered synthesis search. Failing that, \Whiel can produce counterexamples to the validity of Hoare triples, specifically, initial program states from which execution terminates in a state violating the specification. Such counterexamples prove that the program does \emph{not} satisfy the specification.

Consider a developer seeking a view-based rewriting of a \Datalog query. We envision a workflow (Figure~\ref{fig:envisioned-workflow}) in which a coding agent, \LLMsolve, proposes a Datalog program over the view schema. Then, the equivalence of this rewriting to the original query is encoded as an instance of the \Whiel Hoare validity problem, which is passed into the \Whiel system. The \Whiel system utilizes a powerful \emph{guess-and-check} methodology~(Section~\ref{sec:implementation}), with another agent, \LLMwhiel\footnote{We refer to this agent as \AgentHoudini in Figure~\ref{fig:system-overview} and Section~\ref{sec:implementation}.}, generating candidate loop invariants or candidate counterexamples that are then checked by \Whiel's symbolic apparatus. \Whiel outputs a proof if it confirms that \LLMsolve's rewriting is correct; if it returns a counterexample, then \LLMsolve~can use it to revise its attempted rewrite.

\input{figures/envisioned-workflow}

\Whiel (Figure~\ref{fig:system-overview}) combines a flexible \emph{proposer}, an automated symbolic \emph{verifier}, and an independent, LLM-free \emph{certifier}. During iterative search, the verifier checks proposer-provided candidate invariants or counterexamples using our \LH algorithm, a variant of the classical \Houdini algorithm~\cite{flanagan2001houdini}. This algorithm accepts sets of first-order expressible assertions which may contain {prophecy variables} (Definition~\ref{def:prophecy-schema}) representing the terminal values of existing program variables. These assertions are automatically organized by \LH into \emph{leveled assertions} using a pruning process involving \Vampire~\cite{kovacs2013vampire} calls and \Lean-based operations. The verifier provides feedback for further proposals until a proposal is accepted or the resource budget is exhausted. Properties of terminal values established at lower levels support proofs of further properties at higher levels, while the resulting \emph{verification conditions} remain first-order.

The final component of the \Whiel system is the \emph{certifier}, which creates, without LLM-assistance, \Lean certificates that a given Hoare triple is valid or invalid. Invariants are certified using our Lean library combined with Vampire proofs reconstructed via \VampLean~\cite{bodingbauer2026vamplean}, while counterexamples are certified by a relational program interpreter and model-checker implemented in \Lean. Depending on the application, certification is an optional step which reduces the \emph{trusted computing base} of the answer to only the \Lean kernel, removing the need to trust the LLM proposer, verifier implementation, or even \Vampire.

Our experiments (Section~\ref{sec:experiments}) demonstrate the value and power of \Whiel's neurosymbolic architecture. With GPT-5.5, \Whiel's iterative search process is able to solve 73 of the 86 tasks in our curated benchmark. In contrast, our two separate baseline methods, one which attempts to directly generate a \Lean proof without help from our symbolic harness (\directlean), and another which makes only one proposal but does not receive feedback or iterate (\whieloneround), solve 20 and 46 benchmark examples, respectively. In addition to the success rate, \Whiel solves more tasks at an average token cost of \$0.49, compared with \$2.08 for \directlean. The first comparison demonstrates the value of the symbolic harness; the second demonstrates the further benefit of iterative search. \Whiel's 600-second budget covers search and runtime verification; certification is measured separately, since it occurs after an answer has already been accepted by the verifier.

Beyond data management, \Whiel's certified verification can enable reliable LLM-assisted development of relational programs, whose application areas continue to grow: these include declarative networking and network analysis~\cite{loo2009declarative,Zhang2022DiffNetAnalys,zhang2020IncNetVerif}, static program analysis~\cite{whaley2005using,herbert2016souffle}, and enterprise and knowledge-graph applications through systems such as Rel~\cite{aref2025rel} and Vadalog~\cite{bellomarini2018vadalog}.
Scallop and Lobster apply the paradigm to neurosymbolic programming~\cite{li2023scallop,biberstein2025lobster}, alongside work on declarative machine learning~\cite{Wang2021DlogAggML}.
Differential dataflow~\cite{mcSherry2013differential} and incremental and dynamic graph processing~\cite{Fan2021IncGraph,Fan2022BigGraphs,Besta2019DynGraphs} address efficient evaluation over changing data.

The novel contributions of this paper are:
\begin{enumerate}[leftmargin=*]
\itemsep0em
    \item \textbf{A common verification target.}
    We introduce the \Whiel Hoare validity problem (Section~\ref{sec:hoare-validity}), and explain how to encode database problems, including recursive query equivalence and containment, correctness of evaluation strategies, incremental view maintenance, and view-based query rewriting (Section~\ref{sec:applications}).
    \item \textbf{Automated structuring of ladder invariants}
    (Section~\ref{sec:verification}).
    We introduce \emph{ladder invariants} and their compact \emph{leveled assertion} representations, supporting nested inductive arguments in which properties of a loop's final state help establish further properties. Our \LH algorithm automatically organizes proposed clauses into levels reflecting this inductive structure while keeping verification conditions first-order.
    \item \textbf{An implemented neurosymbolic verification system} (Section~\ref{sec:implementation}).
    We develop \Whiel, combining an agent proposer with a symbolic verifier that checks candidate invariants and counterexamples and provides feedback for iterative search. An independent certifier constructs \Lean proofs of validity or invalidity without further LLM assistance.
    \item \textbf{A certified benchmark and experimental evaluation} (Section~\ref{sec:experiments}).
    We provide a benchmark of 86 relational verification tasks with certified reference answers. Our evaluation measures the benefits of \Whiel's iterative search relative to direct \Lean proof generation and single-proposal baselines, and separately reports certification success, time, and memory usage.
\end{enumerate}

\subsection*{Related work}
Theoretical work on database program verification goes back as far as the work by Casanova and Bernstein~\cite{casanova1980formal} in 1980 who proposed a formal system for reasoning about programs accessing a relational database through data definition and manipulation languages. Berghammer et al.~\cite{berghammer2014automated} verify relational programs over binary relations using manually-supplied invariants, discharging their verification conditions using the automated first-order theorem prover Prover9~\cite{mccune2005prover9}. Itzhaky et al.~\cite{itzhaky2017automated} verify loop-free scripts over a fragment of \SQL using weakest precondition computations, with their resulting verification conditions falling within a decidable fragment of first-order logic (\FOL). Mediator~\cite{wang2018verifying} verifies equivalence between pairs of programs implementing loop-free update transactions and nonrecursive queries, with invariants generated using clause templates and \Houdini-style pruning. Beyond database and relational verification, \LOOPY~\cite{kamath2024leveraging} and the work of Pirzada et al.~\cite{pirzada2024llm} utilize LLMs as candidate invariant generators, with the former using \Houdini-style pruning and the latter using \Vampire to avoid unrolling loops during bounded model-checking.

\VeriEQL~\cite{yang2024verieql} proves \emph{bounded} equivalence of \SQL queries. Symbolic tools for proving unbounded equivalence of \SQL queries include \SQLSolver~\cite{ding2023proving}, \QED~\cite{wang2024qed}, and \Cosette~\cite{chu2017cosette}, with the latter even providing proofs verified in the proof assistant \Coq (now \Rocq~\cite{rocq2025prover}). \GraphQE~\cite{tang2025proving} proves equivalence for \Cypher~\cite{francis2018cypher} queries, while \Graphiti~\cite{yang2025graphiti} proves equivalence between \Cypher and \SQL queries. Even more recent work studies verified LLM-guided \SQL rewrites~\cite{lee2026verified}, while a recent preprint introduces \Logos~\cite{ke2026logos}, a tool for verifying \SQL rewrites in \Rocq.

Prophecy variables support reasoning about future program behavior~\cite{lamport2022prophecy, jung2020future}. \Whiel's prophecy termination condition is reminiscent of the use of prophecy variables to establish a property whenever predicted values agree with actual program values~\cite{mcmillan2018eager, mann2022counterexample}. We introduce the \LH algorithm to automatically organize candidate clauses into stages of a nested inductive argument. Within each stage, our algorithm applies the annotation-pruning framework of Flanagan et al.~\cite{flanagan2001houdini} to identify mutually-supporting clauses under assumptions from earlier stages.

%% file: figures/system-overview.tex
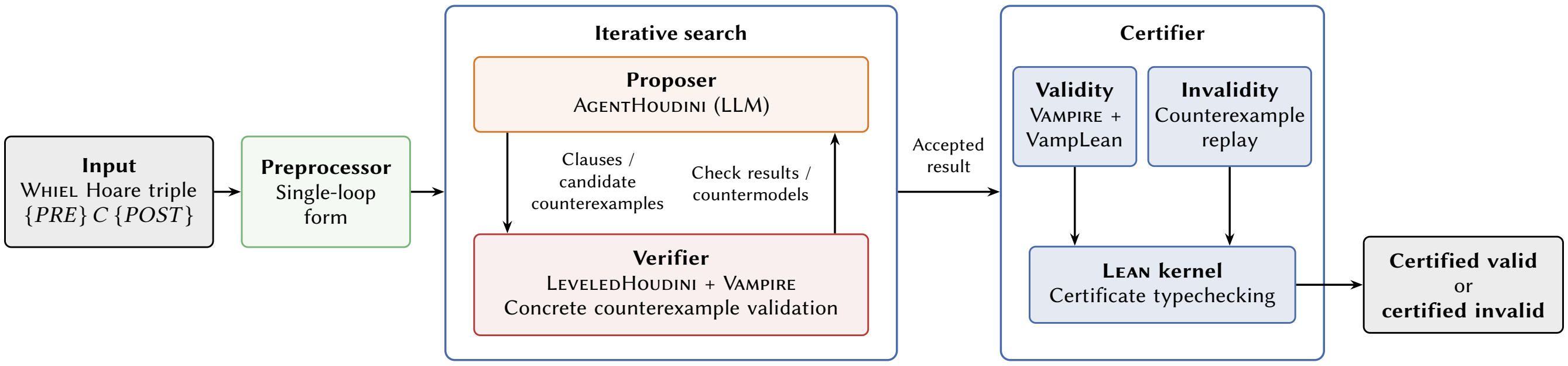
\begin{figure*}[t]
\centering
\begingroup
\definecolor{overviewInput}{gray}{0.85}
\definecolor{overviewPreproc}{RGB}{120,180,120}
\definecolor{overviewSynth}{RGB}{72,106,170}
\definecolor{overviewNeuro}{RGB}{210,120,40}
\definecolor{overviewCheck}{RGB}{180,60,60}
\definecolor{overviewCert}{RGB}{72,106,170}
\color{black}
\begin{tikzpicture}[
    font=\sffamily\footnotesize,
    text=black,
    box/.style={draw=black, rounded corners=2pt, fill=white,
        align=center, inner sep=5pt, line width=.55pt},
    endpoint/.style={fill=overviewInput!50},
    preprocessor box/.style={draw=overviewPreproc, fill=overviewPreproc!8},
    proposer box/.style={draw=overviewNeuro, fill=overviewNeuro!8},
    verifier box/.style={draw=overviewCheck, fill=overviewCheck!8},
    certifier box/.style={draw=overviewCert, fill=overviewCert!15},
    group/.style={draw=black, rounded corners=3pt,
        line width=.65pt, inner sep=0pt},
    flow/.style={-{Stealth[length=4pt,width=3pt]},
        draw=black, line width=.65pt},
    note/.style={font=\sffamily\scriptsize, align=center},
    title/.style={font=\sffamily\bfseries\footnotesize}
]
\node[box, endpoint, text width=2.0cm, minimum height=1.25cm] (input) at (0.8,1.4)
    {\textbf{Input}\\\Whiel Hoare triple\\$\Hoare{PRE}{C}{POST}$};
\node[box, preprocessor box, text width=1.55cm, minimum height=1.25cm] (preprocess) at (3.25,1.4)
    {\textbf{Preprocessor}\\\mbox{Single-loop}\\form};
\draw[flow] (input.east) -- (preprocess.west);

\node[group, draw=overviewSynth, minimum width=5.1cm, minimum height=4.0cm] (search) at (7.15,1.5) {};
\node[title] at (7.15,3.2) {Iterative search};
\node[box, proposer box, text width=4.1cm, minimum height=.68cm] (proposer) at (7.15,2.5)
    {\textbf{Proposer}\\
    \AgentHoudini (LLM)};
\node[box, verifier box, text width=4.1cm, minimum height=1.1cm] (verifier) at (7.15,.35)
    {\textbf{Verifier}\\ \LH + \Vampire\\Concrete counterexample validation};
\draw[flow] ([xshift=-1.85cm]proposer.south)
    -- ([xshift=-1.85cm]verifier.north);
\node[note, text width=1.65cm] at (6.32,1.5)
    {Clauses / candidate\\counterexamples};
\draw[flow] ([xshift=1.85cm]verifier.north)
    -- ([xshift=1.85cm]proposer.south);
\node[note, text width=1.5cm] at (8.05,1.5)
    {Check results /\\countermodels};
\draw[flow] (preprocess.east) -- (preprocess.east -| search.west);

\node[group, draw=overviewCert, minimum width=3.65cm, minimum height=4.0cm] (certifier) at (12.7,1.5) {};
\node[title] at (12.7,3.2) {Certifier};
\node[box, certifier box, inner sep=2pt, text width=1.25cm, minimum height=1.12cm] (proofs) at (11.71,2.25)
    {\textbf{Validity}\\\Vampire + \\VampLean};
\node[box, certifier box, inner sep=2pt, text width=1.7cm, minimum height=1.12cm] (execution) at (13.46,2.25)
    {\textbf{Invalidity}\\Counterexample replay};
\node[box, certifier box, text width=2.65cm, minimum height=.72cm] (kernel) at (12.7,.35)
    {\textbf{\Lean kernel}\\Certificate typechecking};
\draw[flow] (proofs.south) -- (proofs.south |- kernel.north);
\draw[flow] (execution.south) -- (execution.south |- kernel.north);
\draw[flow] (search.east |- preprocess.east) -- (certifier.west |- preprocess.east)
    node[midway, above=3pt, note, text width=1.05cm] {Accepted\\result};

\node[box, endpoint, text width=1.9cm, minimum height=.85cm] (output) at (16.1,.35)
    {\textbf{Certified valid}\\or\\\textbf{certified invalid}};
\draw[flow] (kernel.east) -- (output.west);
\end{tikzpicture}
\endgroup
\caption{Overview of the \Whiel system. Search seeks an invariant satisfying the ladder rule or a concrete counterexample. The certifier produces a \Lean-checked proof of validity or invalidity for the original input triple.}
\label{fig:system-overview}
\Description{An input Hoare triple is preprocessed into single-loop form. A proposer and verifier exchange candidate clauses or counterexamples and check feedback. Accepted results proceed through proof reconstruction or execution certification and Lean kernel checking to a certified valid or invalid result.}
\end{figure*}

%% file: figures/envisioned-workflow.tex
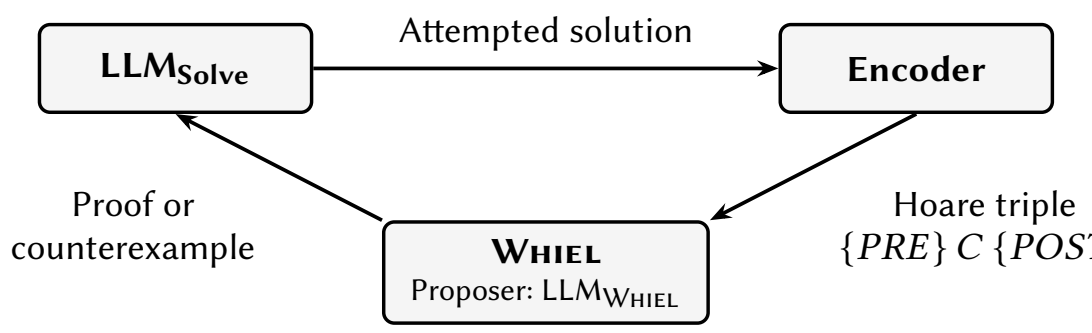
\begin{figure}[htbp]
\centering
\begingroup
\definecolor{workflowBox}{gray}{0.96}
\color{black}
\begin{tikzpicture}[
    font=\sffamily\footnotesize,
    text=black,
    component/.style={rounded corners=2pt, line width=.65pt,
        minimum width=1.8cm, minimum height=.58cm,
        inner xsep=5pt, inner ysep=3pt, align=center},
    flow/.style={-{Stealth[length=4pt,width=3pt]},
        draw=black, line width=.7pt},
    flow label/.style={font=\sffamily\footnotesize,
        align=center, inner sep=2pt}
]
\node[component, draw=black, fill=workflowBox]
    (solve) at (0,0) {\textbf{\LLMsolve}};
\node[component, draw=black, fill=workflowBox]
    (encoder) at (4.9,0) {\textbf{Encoder}};
\node[component, draw=black, fill=workflowBox]
    (whiel) at (2.45,-1.35)
    {\textbf{\Whiel}\\[-1pt]{\scriptsize Proposer: \LLMwhiel}};
\draw[flow] (solve.east) -- (encoder.west)
    node[midway, above=1pt, flow label] {Attempted solution};
\draw[flow] (encoder.south) -- (whiel.north east)
    node[midway, below right=3pt, flow label]
    {Hoare triple\\$\Hoare{PRE}{C}{POST}$};
\draw[flow] (whiel.north west) -- (solve.south)
    node[midway, below left=3pt, flow label]
    {Proof or\\counterexample};
\end{tikzpicture}
\endgroup
\caption{Envisioned workflow for solving a hard database problem with feedback from \Whiel.}
\label{fig:envisioned-workflow}
\Description{Three boxes form a directed triangle. LLM Solve sends an attempted solution to an encoder, which sends a Hoare triple to Whiel. Whiel includes LLM Whiel as its proposer and returns a proof or counterexample to LLM Solve.}
\end{figure}

%% file: sections/s2-hoare-validity.tex
\section{The \Whiel Hoare Validity Problem}
\label{sec:hoare-validity}

\subsection{Preliminaries}

\subsubsection*{Database terminology}
We write $\Gamma$ to denote database schemas, which are finite collections of relation symbols, each with an assigned natural number \emph{arity}. A relation symbol is \emph{nullary} if it has arity $0$. A database instance over $\Gamma$ is a map from the symbols of $\Gamma$ to finite relations of appropriate arity over a fixed data domain.

\emph{Datalog} is a declarative query language. \Datalog \emph{programs} are written as sets of rules. Rules are evaluated by matching the right-hand side of the rule to a database instance, and then deriving the tuple on the left-hand side. Evaluation repeatedly applies these rules until no new tuples can be derived. As an example, the following program computes the transitive closure $T$ of a binary relation $E$:
\input{figures/tcdatalog}

\subsubsection*{Computability}
A decision problem is \emph{recursively enumerable} (\RE) if there is an algorithm which enumerates its positive instances. The class $\coRE$ consists of problems whose complements are in \RE. A problem is complete for a class if it belongs to that class and if any problem in the class can be reduced to it via \emph{many-one reductions}; that is, computable maps that preserve the answers. Many hard decision problems in databases, like those mentioned in the introduction, are \coRE-complete. Consequently, no sound, effective procedure can enumerate all positive instances. An \emph{oracle} for a decision problem is a function that returns the correct answer for all inputs to the problem. An oracle is a construct for theoretical analysis; the function need not be computable.

\subsection{\Whiel Programs and Hoare Triples}
\label{sec:syntax-semantics}

\Whiel \emph{commands} transform database instances via assignments to relation-valued variables over a fixed \emph{execution schema}. \emph{States} for such commands are database instances over the execution schema; that is, states are maps from the schema symbols to finite relations of appropriate arity. Assignments update variables using \RA expressions; guards of \while loops and \ite conditionals are \Whiel \emph{assertions}: Boolean combinations of inclusions and equalities between \RA expressions. 
The formal syntax is given by
\input{figures/syntax.tex}

Here, $i$ and $j$ denote tuple indices, $\varphi$ denotes selection conditions, $R$ and $X$ are relation symbols, $d$ denotes a domain constant, $E$ denotes \RA expressions, $A$ denotes \Whiel assertions, and $C$ denotes \Whiel commands. \RA expressions are evaluated in states (i.e., database instances) as usual, while \emph{assertions} are either satisfied or falsified. We write $E(s)$ to denote the result of evaluating \RA expression $E$ in state $s$. For assertions, $E_1 \subseteq E_2$ is satisfied in $s$ (notation: $s \models E_1 \subseteq E_2$) if $E_1(s) \subseteq E_2(s)$, and similarly for equalities. The semantics of assertions are extended to Boolean combinations in the natural way. Given a state $s$, we write $s \models A$ if $A$ evaluates to true in $s$ (under the natural semantics for the guard syntax), in which case we say that $A$ is \emph{satisfied} in $s$.

Assertion $A$ \emph{entails} assertion $B$ (notation: $A \models B$) if every state satisfying $A$ also satisfies $B$; in this case we say that the \emph{entailment} $A \models B$ is \emph{valid}. Using \emph{active domain semantics}~\cite{abiteboul1995foundations}, Codd's theorem~\cite{codd1970relational} implies that \Whiel assertions can be translated to sentences of first-order logic (\FOL), and that every sentence of \FOL is expressed by some \Whiel assertion. \Whiel includes a Lean-formalized translation from assertions to first-order logic, enabling the system to check entailments between assertions using \Vampire.

\Whiel is endowed in \Lean with a formalized big-step semantics, mapping input states to output states. Assignments update assigned variables to the result of evaluating \RA expressions; $C_1 \seq C_2$ executes $C_1$ followed by $C_2$; \sk does nothing; and \while and \ite have their usual guarded semantics. The semantics are partial (i.e., the output state is sometimes undefined) due to the potential for nonterminating \while loops.

The transitive closure \Datalog program~\ref{eq:tcdatalog} can be simulated in \Whiel (see Program~\ref{eq:tcwhiel} below). Borrowing terminology from \Datalog, we refer to relational variables which appear on the left-hand side of some assignment as \emph{intensional database predicates} (IDBs), and all other relational variables as \emph{extensional database predicates}. Given a \Whiel command $C$, we write $\IDB(C)$ (resp. $\EDB(C)$) to denote its intensional (resp. extensional) database predicates. For example, in the program below, we have that $\IDB(\tcwhiel) = \{ S, T \}$, while $\EDB(\tcwhiel) = \{ E \}$.

\input{figures/tcwhiel.tex}

We write $E \circ T$ as shorthand for relational composition, which is definable in \RA using a select-project-join query. Note that \tcwhiel uses an \emph{auxiliary variable}\footnote{A variable which is not part of the ``intended'' output, but supports the computation.} $S$ to capture the next value for $T$. Under appropriate assumptions on the initialization of $S$ and $T$ (see Equation~\ref{eq:tcwhiel-triple}), the program computes the transitive closure of $E$. Auxiliary variables are not strictly necessary for simulating \Datalog programs in \Whiel, but we include $S$ for expository purposes.

Inputs to the \Whiel system are \emph{Hoare triples}, written
\begin{equation}
\label{eq:hoare-triple}
    \Hoare{PRE}{C}{POST},
\end{equation}
where $PRE$ and $POST$ are \emph{assertions} and $C$ is a \Whiel command. A Hoare triple is \emph{valid} if, for all terminating runs of $C$ starting from a state satisfying $PRE$, the halting state satisfies $POST$. The triple is \emph{invalid} otherwise. Thus, providing a single state which satisfies $PRE$ and for which $C$ terminates in a state falsifying $POST$ suffices to prove the triple invalid. The goal of the \Whiel system is to determine whether or not a given Hoare triple is valid; this is the \emph{\Whiel Hoare validity problem}.

\begin{theorem}
    \label{thm:whv-coRE-complete}
    \Whiel Hoare validity is \coRE-complete.
\end{theorem}

As an example, consider the following Hoare triple:
\begin{equation}
    \label{eq:tcwhiel-triple}
    \Hoare{T = \emptyset \land S = E}{\quad\tcwhiel\quad}{T \circ T \subseteq T}.
\end{equation}
This asserts that, assuming that the IDB $T$ is initially empty while $S$ is initialized to $E$, the final value of $T$ after running the $\tcwhiel$ program is transitive.

%% file: figures/tcdatalog.tex
\begin{equation}
\tag{\textbf{TC$_\Datalog$}}
\label{eq:tcdatalog}
\begin{aligned}
    T(x, y) &\dlog E(x, y) \\
    T(x, y) &\dlog E(x, z),~ T(z, y).
\end{aligned}
\end{equation}

%% file: figures/syntax.tex
\[
\begin{aligned}
    \varphi &::= i=j \mid i=d \mid \lnot \varphi \mid \varphi \land \varphi \mid \varphi \lor \varphi \\
    E &::= \top \mid \emptyset \mid R \mid \{d\} \mid \select_{\varphi} E \mid \project_{L} E \mid E \times E \mid E \cup E \mid E - E \\
    A &::= \true \mid \false \mid E \subseteq E \mid E = E \mid \lnot A \mid A \land A \mid A \lor A \\
    C &::= X := E \mid C \seq C \mid \sk \mid \iteargs{A}{C}{C} \mid \whileargs{A}{C}
\end{aligned}
\]

%% file: figures/tcwhiel.tex
\begin{equation}
\tag{\tcwhiel}
\label{eq:tcwhiel}
    \begin{array}{@{}l@{}}
    \while~T \neq S~\mathbf{do} \\
    \quad T := S \\
    \quad S := E \cup E \circ T \\
    \mathbf{end}
    \end{array}
\end{equation}

%% file: sections/s3-applications.tex
\input{figures/red-blue-paths}

\section{Applications of the \Whiel System}
\label{sec:applications}

\Whiel can be applied to hard problems in databases by encoding them as instances of the \Whiel Hoare validity problem. This may initially seem unsurprising, since the problem's \coRE-completeness entails that \emph{any} problem in the class of \coRE, including many database problems, can be reduced to \Whiel Hoare validity. However, it is striking how \emph{natural} and \emph{compact} these encodings are, and how efficiently they can be constructed.

\subsubsection*{Encoding \Datalog}
Consider the \Datalog \emph{path system} query~\cite{cook1974observation}:
\input{figures/path-system}

Here, $A$ denotes a set of initially accessible nodes, while a tuple $R(x,y,z)$ says that node $x$ is accessible if both $y$ and $z$ are accessible. We will use this example to illustrate \Whiel's ability to express different evaluation strategies for \Datalog. Define the shorthand
\begin{equation*}
\label{eq:path-system-step}
    H(X,Y) \equiv
    \project_{[0]} \left( \select_{1=3 \land 2=4}((R\times X)\times Y) \right).
\end{equation*}
The following two programs simulate \emph{naive} and \emph{semi-naive} evaluation~\cite[Section~13.1]{abiteboul1995foundations}, respectively:
\input{figures/path-system-whiel}

\subsubsection*{Encoding Graph Queries}

Property graph query languages like GQL, SQL/PGQ, and Cypher are built on a core of pattern matching and relational operations~\cite{gheerbrant2025gql}. In~\cite{han2024implementation}, the authors encode property graphs into relational database instances, using Datalog to implement query rewriting with views. Borrowing from this encoding, \Whiel can represent pattern matches using relations holding their endpoints and variable bindings. Concatenation is handled by composing matches with shared endpoints and compatible bindings, negated conditions are represented with relational difference, and unbounded repetition uses while loops to accumulate endpoint pairs until a fixed point is reached. After simulating pattern matching, \Whiel can express any \CoreGQL query using relational operations.

Using Datalog instead of graph query syntax, Equation~\ref{eq:answering-datalog-with-datalog-views} illustrates this connection. The query $T$ computes reachability along all edges, red or blue; $\mathit{RedPV}$ and $\mathit{BluPV}$ are views computing the red-only and blue-only paths, respectively. The rewriting $R$ recovers $T$ by composing paths supplied by these views. In general, deciding whether or not an equivalent  \Datalog rewriting using views exists is undecidable~\cite{duschka1997answering}. We have encoded this example in \Whiel using separate binary relations for red and blue edges, and verified and certified the equivalence of the original query and its rewriting. 

More generally, \Whiel can express interesting graph computations
such as computation of the \emph{attractor set} for two-player \emph{reachability games} on finite graphs~\cite{mazala2002infinite}. In such games, we are given a graph $(V,E)$ with no dead ends and subsets of vertices $P_a, Target \subseteq V$. In such a game, there is an \emph{attacker} whose goal is to reach the $Target$ vertices. The goal of the \emph{defender} is to prevent the attacker from reaching the $Target$ set. The attacker may choose the next node from vertices in $P_a$; other nodes belong to the defender. The attractor of the game is the set of nodes $W$ from which the attacker has a strategy to force the game position to reach the $Target$ set, irrespective of the moves chosen by the defender. Define $\mathrm{Pred}(X) := \project_{[0]}(\select_{1=2}(E \times X))$.
The following monotone operator on the powerset of vertices has, as its least fixed point, the \emph{attractor set} of the graph to which it is applied:
\input{figures/reachability-game-step}
Using this operator, the corresponding \Whiel command is
\input{figures/reachability-game-whiel}

Several interesting correctness properties of this command can be expressed in \Whiel's assertion language. Let $\Sigma$ denote some collection of first-order constraints on the input graph, and let $I$ be a subset of vertices. To check whether these constraints guarantee that $I$ is \emph{safe}, meaning that the attacker cannot win from any vertex in $I$, we can ask whether the following Hoare triple is valid:
\input{figures/reachability-game-triple}

\subsubsection*{Encoding Other Database Problems}
Using difference, \Whiel can also simulate variants of \Datalog with negation. Our assertion language can encode first-order constraints (such as tuple- or equality-generating dependencies) in the precondition, and compare the corresponding output relations by equality or inclusion, allowing \Whiel to encode containment and equality problems subject to constraints. \Whiel can also express the correctness of incremental view computation algorithms.
As an example, the following program computes the updated output $T'$ from the output $T$ of \psdatalog after an insertion of nodes $\Delta$ into $A$.
\begin{equation}
\tag{\ivmwhiel}
\label{eq:ivm-whiel}
\begin{array}{@{}l@{}}
A' := A \cup \Delta
    \seq T':=T
    \seq D := \Delta \setminus T' \\
\whileargs{D \neq \emptyset}{\\
    \quad T':=T'\cup D\\
    \quad D:=\bigl(H(D,T')\cup H(T',D)\bigr)\setminus T'\\
}
\end{array}
\end{equation}
This computation is correct if the \Whiel program sequencing \psnaive, \ivmwhiel, setting $A := A'$, and then re-running \psnaive yields $T = T'$ on all inputs, which is expressible as a Hoare triple. Schema mappings can also be expressed as first-order constraints. For view-based rewritings, we first compute the view, and then compare the result of the original query with the result of the rewritten query applied to the view schema.

%% file: figures/red-blue-paths.tex
\begin{figure*}[t]
\centering
\setlength{\columnwidth}{\textwidth}
\setlength{\belowdisplayskip}{0pt}
\setlength{\belowdisplayshortskip}{0pt}
\begin{equation}
\label{eq:answering-datalog-with-datalog-views}
\begin{array}{@{}l@{\qquad}l@{\qquad}l@{}}
\text{\bfseries Query $T$} & \text{\bfseries Views} & \text{\bfseries Rewriting $R$} \\[4pt]
\begin{aligned}[t]
T(x,y)&\dlog E(x,\mathit{red},y)\\
T(x,y)&\dlog E(x,\mathit{blu},y)\\
T(x,z)&\dlog E(x,\mathit{red},y), T(y,z)\\
T(x,z)&\dlog E(x,\mathit{blu},y), T(y,z)
\end{aligned}
&
\begin{aligned}[t]
\mathit{RedPV}(x,y)&\dlog E(x,\mathit{red},y)\\
\mathit{RedPV}(x,z)&\dlog E(x,\mathit{red},y), \mathit{RedPV}(y,z)\\[4pt]
\mathit{BluPV}(x,y)&\dlog E(x,\mathit{blu},y)\\
\mathit{BluPV}(x,z)&\dlog E(x,\mathit{blu},y), \mathit{BluPV}(y,z)
\end{aligned}
&
\begin{aligned}[t]
R(x,y)&\dlog \mathit{RedPV}(x,y)\\
R(x,y)&\dlog \mathit{BluPV}(x,y)\\
R(x,z)&\dlog R(x,y),R(y,z)
\end{aligned}
\end{array}
\vspace*{-1em}
\end{equation}
\Description{Datalog rules in three groups: the query T for red-or-blue paths, red and blue path views, and the rewriting R using those views.}
\end{figure*}

%% file: figures/path-system.tex
\begin{equation}
\tag{\psdatalog}
\label{eq:path-system}
\begin{aligned}
    T(x)&\dlog A(x),\\
    T(x)&\dlog R(x,y,z),\ T(y),\ T(z).
\end{aligned}
\end{equation}

%% file: figures/path-system-whiel.tex
\begin{equation*}
\label{eq:path-system-whiel}
    \begin{array}{@{}l@{\quad\quad}l@{}}
    \psnaive&\psseminaive\\[2pt]
    \begin{array}[t]{@{}l@{}}
    T:=A\\
    S:=A\cup H(T,T)\\
    \mathbf{while}\ T\ne S\ \mathbf{do}\\
    \quad T:=S\\
    \quad S:=A\cup H(T,T)\\
    \mathbf{end}
    \end{array}
    &
    \begin{array}[t]{@{}l@{}}
    U:=A\seq\Delta:=A\\
    \mathbf{while}\ \Delta\ne\emptyset\ \mathbf{do}\\
    \quad N:=\bigl(H(\Delta,U) \cup H(U,\Delta)\bigr)\setminus U\\
    \quad U:=U\cup N\\
    \quad\Delta:=N\\
    \mathbf{end}
    \end{array}
    \end{array}
\end{equation*}

%% file: figures/reachability-game-step.tex
\begin{equation*}
\label{eq:reachability-game-step}
    F(X) = Target \cup ( P_a \cap \mathrm{Pred}(X)) \cup \left( (V \setminus P_a) \setminus \mathrm{Pred}(V \setminus X) \right).
\end{equation*}

%% file: figures/reachability-game-whiel.tex
\begin{equation}
\tag{\rgwhiel}
\label{eq:reachability-game-whiel}
\begin{array}{@{}l@{}}
W:=\emptyset\seq Z:=F(W)\\
\mathbf{while}\ W\ne Z\ \mathbf{do}\\
\quad W:=Z\\
\quad Z:=F(W)\\
\mathbf{end}
\end{array}
\end{equation}

%% file: figures/reachability-game-triple.tex
\begin{equation*}
\label{eq:reachability-game-triple}
    \Hoare{\Sigma}{\quad\rgwhiel\quad}{I \cap W = \emptyset}.
\end{equation*}

%% file: sections/s4-verification.tex
\section{\Whiel's Verifier}
\label{sec:verification}

In this section, we explain the mathematical theory behind \Whiel's verifier. For this, we return to our expository example, \tcwhiel.

\subsection{Hoare Logic and Verification Conditions}
\label{sec:hoare}

The key challenge in solving the \Whiel Hoare validity problem is the identification of \emph{loop invariants}; that is, given a Hoare triple
\begin{equation}
\label{eq:loop-triple}
    \Hoare{PRE}{\whileargs{G}{B}}{POST},
\end{equation}
to identify an assertion $I$ satisfying the \emph{verification conditions}
\begin{gather*}
    PRE \models I,
        \tag*{$\init_I$} \\
    \Hoare{I \land G}{B}{I}~\text{is valid, and}
        \tag*{$\maint_I$} \\
    I \land \lnot G \models POST.
        \tag*{$\term_I$}
\end{gather*}
We refer to these as the \emph{initialization}, \emph{maintenance}, and \emph{termination} conditions, respectively. Note that $\init_I$, $\maint_I$, and $\term_I$ depend on $PRE$, $POST$, $G$, and $B$ from the Hoare triple in Equation \ref{eq:loop-triple}. We leave this dependence on the Hoare triple implicit. We call $I$ \emph{inductive} if $\init_I$ and $\maint_I$ hold. Intuitively, these three conditions assert that the assertion $I$ is satisfied at the start of the loop ($\init_I$), after each loop iteration ($\maint_I$), and that $I$ together with the negation of the loop guard implies $POST$ ($\term_I$).

\begin{proposition}
    \label{prop:inv-rule}
    If there exists an assertion $I$ satisfying $\init_I$, $\maint_I$, and $\term_I$, then the Hoare triple in Equation~\ref{eq:loop-triple} is valid.
\end{proposition}

Since $PRE$, $POST$, $I$, and $G$ can be translated to \FOL, the $\init_I$ and $\term_I$ checks are effectively \FOL entailments. To express $\maint_I$ as an \FOL entailment, we use the \emph{weakest liberal precondition of $I$ with respect to $B$}, which is the set of states from which all terminating runs of $B$ halt in a state satisfying $I$~\cite{dijkstra1976discipline}. This set is \emph{expressible} if there exists a \Whiel assertion $\wp(I,B)$ which is satisfied in all and only these states.
\begin{proposition}
\label{prop:maint-wp}
    For an assertion $I$ and loop-free\footnote{that is, containing no occurrences of \while} command $B$, $\wp(I,B)$ exists and is computable from $I$ and $B$; and for every \Whiel assertion $G$, $\Hoare{I \land G}{B}{I}$ is valid if and only if $I \land G \models \wp(I,B)$.
\end{proposition}

For an assignment $X := E$, we have that
$\wp(A, X := E) \equiv A[ E / X ]$, where $A[ E / X ]$ denotes the result of simultaneously replacing occurrences of $X$ in $A$ with $E$. For a conditional with guard $H$ and loop-free branches $C_1$ and $C_2$,
\[
\begin{aligned}
    &\wp(A, \iteargs{H}{C_1}{C_2}) \\
    &\quad\quad\equiv (H \to \wp(A, C_1)) \land (\lnot H \to \wp(A, C_2)).
\end{aligned}
\]
Intuitively, the $\wp$ rule for \ite asserts that the precondition of the selected branch must hold. Skip is straightforward, with $\wp(A,\sk)\equiv A$, while sequencing iterates \wp computation backward from the last command in the sequence.

Thus, when $B$ is loop-free, all three verification conditions for Equation~\ref{eq:loop-triple} can be written as \FOL entailments. For the remainder of this section, we consider commands in \emph{single-loop form}; that is, of the form in Equation~\ref{eq:loop-triple}, where $B$ is loop-free.

In practice, it is convenient to search for \emph{conjunctive invariants} $I$, which are invariants given as the conjunction of some set of smaller assertions, called \emph{clauses}\footnote{Clauses are often assumed to have a particular shape, such as being disjunctions of literals, but the \Whiel system uses ``clause'' and ``assertion'' interchangeably.}. When an invariant is written as a conjunction of clauses, the verification conditions $\init_I$ and $\maint_I$ can be decomposed into \emph{per-clause} checks:
\begin{align*}
    PRE &\models c,~\text{and}
        \tag*{$\init_c$} \\
    I \land G &\models \wp(c,B).
        \tag*{$\step_{I,c}$}
\end{align*}
We refer to these as the \emph{initialization} and \emph{preservation} conditions, respectively. Observe that in the definition of $\step_{I,c}$, the entire invariant $I$ is available on the left-hand side of the entailment. A clause $c$ is \emph{preserved} by $I$ if $\step_{I,c}$ holds. The following proposition justifies this decomposition.
\begin{proposition}
\label{prop:init-maint-per-clause}
    Let $I = \bigwedge_{c \in S} c$ denote the conjunction of a finite set $S$ of clauses. Then
    \begin{enumerate}
        \item $\init_I$ holds if and only if $\init_{c}$ holds for all $c \in S$; and
        \item $\maint_I$ holds if and only if $\step_{I,c}$ holds for all $c \in S$.
    \end{enumerate}
\end{proposition}
The \emph{Houdini algorithm}~\cite{flanagan2001houdini} uses this decomposition to compute the \emph{maximal inductive subset}\footnote{This guarantee presupposes that the verification condition checks are conclusive.} of a finite set $S$ of clauses. That is, it computes the largest subset $S' \subseteq S$ such that $\init_{I'}$ and $\maint_{I'}$ hold, where $I' = \bigwedge_{s \in S'} s$. If the resulting invariant $I'$ also satisfies $\term_{I'}$, then the Hoare triple is valid. \Whiel employs a modified version of the Houdini algorithm (Section~\ref{sec:leveled-houdini}) which is specialized to handle the \emph{prophecy variables} and \emph{leveled invariants} described in Sections~\ref{sec:prophecy} and ~\ref{sec:leveled-invariants}, respectively.

\subsection[Prophecy Variables]{Prophecy Variables}
\label{sec:prophecy}

Consider again the triple in Equation~\ref{eq:tcwhiel-triple}. For this running example, we write $G$ and $B$ to denote the guard $T \neq S$ and the body $T := S \seq S := E \cup E \circ T$ of \tcwhiel, respectively. One initial attempt at providing an invariant is to use $c_0 \equiv S = E \cup E \circ T$, which is easily seen to satisfy $\init_{c_0}$ and $\maint_{c_0}$. However, termination fails:
\[
    (S = E \cup E \circ T) \land S = T \quad\not\models\quad T \circ T \subseteq T
\]
is an invalid entailment. This is because $c_0$ and $S=T$ imply that $T$ is a fixed point of the map $X \mapsto E \cup (E \circ X)$, but do not guarantee that $T$ is transitive. Furthermore, the assertion $T \circ T \subseteq T$ itself is too strong to be an invariant; while it is initially true (since $T = \emptyset$), and it clearly satisfies the termination condition, it fails to be preserved in general during executions of the loop.

\subsubsection*{Prophecy Variables and Terminal Consequences}
We conjecture that no invariant expressible in \FOL over the execution schema $\{ E, S, T \}$ exists for the Hoare triple in Equation~\ref{eq:tcwhiel-triple}. This seems to be related to the fact that \FOL cannot express transitive closure: the postcondition concerns only the terminating value of $T$, which is precisely the transitive closure of $E$. To mitigate this apparent lack of expressivity, we introduce the following definitions.

\begin{definition}[Prophecy variables and schema]
\label{def:prophecy-schema}
Let $C$ be a \Whiel command over execution schema $\Gamma$. For each $P\in\IDB(C)$, introduce a fresh relation symbol $P^\infty\notin\Gamma$ of the same arity as $P$, called a
\emph{prophecy variable}. The \emph{prophecy schema} of $C$ is
\[
    \Gamma^\infty=\Gamma\cup\{P^\infty\mid P\in\IDB(C)\}.
\]
\end{definition}

The intended interpretation of $P^\infty$ is that it denotes the \emph{terminal value} of $P$, its value at the end of a terminating run of the program $C$. Note that EDBs are not assigned prophecy variants because, by definition, their values are unchanged during the execution of $C$.

\begin{definition}[Prophecy variants and terminal consequences]
\label{def:prophecy-variants}
Let $C$ be a \Whiel command over execution schema $\Gamma$. For a relation symbol $P\in\IDB(C)$, its \emph{prophecy variant} is the corresponding prophecy variable $P^\infty$. For an assertion $A$ over $\Gamma^\infty$, its \emph{terminal consequence} $A^\infty$ is obtained by simultaneously replacing each occurrence of every $P\in\IDB(C)$ with $P^\infty$. EDB symbols and prophecy variables already present in $A$ are unchanged.
\end{definition}

\Whiel supports invariants over the prophecy schema of its inputs. Under the intended interpretation of prophecy variables, $A^\infty$ expresses what $A$, if inductive, ``says'' about the terminal state. However, this alone does not increase the expressive power of the system; any assertion over $\Gamma^\infty$ satisfying the three verification conditions $\init_I$, $\maint_I$, and $\term_I$ can be transformed into an invariant over the original schema $\Gamma$. To strengthen our system, we must modify the verification conditions themselves. The first way to do this is to hard-code the intended meaning of the prophecy variables in the termination condition. Keeping $\init_I$ and $\maint_I$ unchanged, we can replace the $\term_I$ check with
\[
    I \land \lnot G \land \bigwedge_{P \in \IDB(C)} P = P^\infty \models POST.
\]
Equivalently, we can check
\begin{equation}
    \label{eq:term-prophecy}
    I^\infty \land \lnot G^\infty \models POST^\infty.
    \tag*{$\term^\infty_I$}
\end{equation}
The program does not contain occurrences of the prophecy symbols, and so the entailments $\init_I$ and $\maint_I$ must hold for \emph{arbitrary relations} assigned to the prophecy variables; their intended meaning is applied only in the termination check.

\begin{definition}[Prophecy invariant]
    A \emph{prophecy invariant} for a Hoare triple of the form in Equation~\ref{eq:loop-triple}, where $C$ is a \Whiel program over execution schema $\Gamma$, is an assertion $I$ over $\Gamma^\infty$ such that $\init_I$, $\maint_I$, and $\term^\infty_I$ are valid.
\end{definition}

\begin{proposition}[Prophecy rule]
\label{prop:prophecy-rule}
    If a prophecy invariant exists for a Hoare triple of the form in
    Equation~\ref{eq:loop-triple}, then the triple is valid.
\end{proposition}

\subsection{Leveled Invariants}
\label{sec:leveled-invariants}

Proposition~\ref{prop:prophecy-rule} enables us to construct an invariant for Equation~\ref{eq:tcwhiel-triple}. Consider the assertion $c_1 \equiv T \circ T^\infty \subseteq T^\infty$, which, unlike $T \circ T \subseteq T$, is true at the start of the $\tcwhiel$ loop and holds on every iteration (under the intended interpretation of $T^\infty$). Furthermore, we have that $c^\infty_1 \equiv POST^\infty \equiv T^\infty \circ T^\infty \subseteq T^\infty$, and therefore any conjunctive invariant $I$ with $c_1$ as a clause suffices to ensure that $\term^\infty_I$ holds. However, more is needed since $\maint_{c_1}$ fails due to the fact that, while $c_1$ holds at the start of each loop iteration \emph{under the intended interpretation}, Proposition~\ref{prop:prophecy-rule} requires that $\maint_I$ holds for \emph{arbitrary} values assigned to $T^\infty$, which is not the case.

We know from our earlier discussion that $c_0$ is inductive, and it gives some information about the terminating values of $S$ and $T$. In particular, its terminal consequence is $c^\infty_0 \equiv S^\infty = E \cup E \circ T^\infty$, while that of the negation of the loop guard, $\lnot G^\infty$, is $T^\infty = S^\infty$.
Together, these entail that $T^\infty = E \cup E \circ T^\infty$, and so
\begin{equation}
\label{eq:tc-exit-closure}
    E \circ T^\infty \subseteq T^\infty.
\end{equation}
This last fact suffices to preserve $c_1$; more precisely, we establish the entailment
\begin{equation}
\label{eq:tc-preservation}
    c_0\land c_1\land\lnot G^\infty\land c_0^\infty
      \models\wp(c_1,B).
\end{equation}
Here, $c_0$ and $c_1$ should be interpreted as assumptions about the current values of the IDBs, while $\lnot G^\infty\land c_0^\infty$ is an assumption about the terminating values. Since $B$ first assigns $T := S$ and then updates only $S$, we have that $\wp(c_1,B) \equiv S\circ T^\infty\subseteq T^\infty$. Using $c_0$, $c_1$, and the derived fact in Equation~\ref{eq:tc-exit-closure}, we obtain through standard reasoning steps that $S \circ T^\infty \subseteq T^\infty$, which establishes the preservation entailment in Equation~\ref{eq:tc-preservation}. In general, a preservation proof may also assume $G$ (that the guard is true for the current values); we did not need to use this assumption in the above.

Intuitively, the argument proceeds in two stages. The first stage establishes that $c_0$ is inductive. The second stage uses the assumption that $c_0$ is inductive, allowing us to assume its terminal consequence $c^\infty_0$. This combines with the assumption $\lnot G^\infty$, asserting that the guard is false for the terminal values, to derive the useful fact in Equation~\ref{eq:tc-exit-closure}. This useful fact allows us to derive $S \circ T^\infty \subseteq T^\infty$, establishing that $c_1$ holds on each iteration of the \tcwhiel loop, \emph{given the terminal consequences $c^\infty_0$ and $\lnot G^\infty$}. Finally, the terminal consequence $c^\infty_1$ suffices to prove the termination condition of Proposition~\ref{prop:prophecy-rule}. This suggests the following prophecy invariant:
\begin{equation}
\label{eq:tc-ladder}
    \Itc \quad\equiv\quad c_0 \land \left( ( \lnot G^\infty \land c^\infty_0)\to c_1 \right).
\end{equation}

\subsubsection*{Leveled verification conditions}

Consider the per-clause checks of Proposition~\ref{prop:init-maint-per-clause} for each top-level conjunct of $\Itc$. For $c_0$, these are
\begin{align}
    PRE &\models c_0, \quad\text{and}
        \tag*{$\init_{c_0}$}
        \label{eq:tc-level0-checks} \\
    \Itc \land G &\models \wp(c_0, B).
        \tag*{$\step_{\Itc,c_0}$}
        \label{eq:tc-level0-step}
\end{align}
For $d_1 \equiv (\lnot G^\infty \land c^\infty_0) \to c_1$, they are
\begin{align*}
    PRE &\models (\lnot G^\infty \land c^\infty_0) \to c_1, \quad\text{and}
    \tag*{$\init_{d_1}$}
    \label{eq:tc-level1-checks} \\
    \Itc \land G &\models \wp \left( ( \lnot G^\infty \land c^\infty_0)\to c_1,B \right).
    \tag*{$\step_{\Itc,d_1}$}
    \label{eq:tc-level1-step}
\end{align*}
Since $\lnot G^\infty$ and $c^\infty_0$ do not contain IDBs, they are unchanged by taking the weakest liberal precondition with respect to $B$:
\begin{align*}
    &\wp \left( ( \lnot G^\infty \land c^\infty_0)\to c_1, B \right)
    \quad\equiv\quad
    ( \lnot G^\infty \land c^\infty_0 ) \to \wp(c_1, B).
\end{align*}
Consequently, we can simplify the checks of the second conjunct of $\Itc$ to the equivalent entailments
\begin{align*}
    PRE \land \lnot G^\infty \land c^\infty_0 &\models c_1, \quad\text{and} \\
    c_0 \land c_1 \land G \land \lnot G^\infty \land c^\infty_0 &\models \wp(c_1, B).
\end{align*}
Our previous argument establishes that~\ref{eq:tc-level0-checks} and~\ref{eq:tc-level0-step}, as well as the above equivalent versions of~\ref{eq:tc-level1-checks} and~\ref{eq:tc-level1-step}, are valid. Thus by Proposition~\ref{prop:init-maint-per-clause}, $\Itc$ is inductive. Furthermore,
\[
    \Itc^\infty
    \quad\equiv\quad
    c^\infty_0 \land \left( ( \lnot G^\infty \land c^\infty_0) \to c^\infty_1 \right),
\]
and we know that $c^\infty_1 \equiv POST^\infty$, so
\begin{equation}
\tag*{$\term^\infty_{\Itc}$}
\Itc^\infty \land \lnot G^\infty \models POST^\infty
\end{equation}
holds. Hence we can conclude by Proposition~\ref{prop:prophecy-rule} that the Hoare triple in Equation~\ref{eq:tcwhiel-triple} is valid.

\subsubsection*{Generalizing the argument}
The form of the prophecy invariant $\Itc$ can be generalized to support similar stage-based reasoning for \emph{arbitrary} Hoare triples of the form in Equation~\ref{eq:loop-triple}, where the command $C$ has execution schema $\Gamma$.

\begin{definition}
\label{def:leveled-candidate}
    A \emph{leveled candidate} is a tuple $\Ascr = (S_0, \hdots, S_m)$ of finite sets of clauses (called \emph{levels}) over the prophecy schema $\Gamma^\infty$, where $S_0$ contains no occurrences of prophecy variables. We write $\bigcup \Ascr = \bigcup_{j=0}^{m} S_j$ to denote the set of all clauses in $\Ascr$. The \emph{ladder assertion} of $\Ascr$ with respect to $C$ is the assertion
    \begin{equation}
    \label{eq:ladder-invariant}
        L_{\Ascr}
        \quad\equiv\quad
        A_0 \land \bigwedge_{k=1}^{m}
        \left( \left( \lnot G^\infty \land \bigwedge_{i<k} A_i^\infty \right)
        \rightarrow A_k \right),
    \end{equation}
    where $A_k = \bigwedge_{c \in S_k} c$ and empty conjunctions denote $\true$.
\end{definition}
The invariant \Itc in Equation~\ref{eq:tc-ladder} is exactly the ladder assertion of the leveled candidate $\Atcl = (\{ c_0 \}, \{ c_1 \})$ with respect to \tcwhiel. In the general case, we can also introduce modified verification conditions which match the argument for \tcwhiel. For the level-$0$ clauses, these are the typical conditions:
\begin{align}
    PRE &\models c, \quad\text{and}
        \tag*{$\init^0_{\Ascr,c}$}
        \label{eq:ladder-level0-vcs} \\
    G\land A_0 &\models\wp(c,B).
        \tag*{$\step^0_{\Ascr,c}$}
        \label{eq:ladder-level0-step}
\end{align}
For $j > 0$, the checks mirror the argument for \tcwhiel. When checking initialization at a level $S_j$ for $j > 0$, we may assume that the loop guard is false for the terminating values, and that all strictly lower levels satisfied their leveled initialization and preservation checks (and hence we may assume their terminal consequences). Thus the \emph{level-$k$ initialization} condition is
\begin{equation}
\label{eq:leveled-init-vc}
    PRE\land\lnot G^\infty\land\bigwedge_{i<k}A_i^\infty
      \models c.
    \tag*{$\init^k_{\Ascr,c}$}
\end{equation}

The preservation check may assume that the guard is true at the current values, that it is false for the terminal values, that clauses at the current level and below are satisfied on the current iterations, and that all strictly lower levels satisfied their leveled initialization and preservation checks. Thus the \emph{level-$k$ preservation} condition is
\begin{equation}
\label{eq:leveled-step-vc}
    G \land \bigwedge_{i \leq k} A_i \land \lnot G^\infty \land \bigwedge_{i<k} A_i^\infty \models \wp(c, B).
    \tag*{$\step^k_{\Ascr,c}$}
\end{equation}
The termination condition $\term_\Ascr$ for a leveled candidate $\Ascr = (S_0, \hdots, S_m)$ is precisely $\term^\infty_{A}$, where $A = \bigwedge_{k \leq m} A_k$.

\begin{definition}
    A leveled candidate $\Ascr = (S_0, \hdots, S_m)$ is an \emph{(inductive) core} if $\init^k_{\Ascr,c}$ and $\step^k_{\Ascr,c}$ are valid for all $k \leq m$ and $c \in S_k$. It is a \emph{leveled invariant} if $\term_\Ascr$ is valid as well.
\end{definition}

\begin{proposition}[Ladder Rule]
\label{prop:ladder-rule}
    If a leveled invariant $\Ascr$ exists for a Hoare triple of the form in Equation~\ref{eq:loop-triple}, then the triple is valid. Furthermore, $L_{\Ascr}$ is a prophecy invariant for the same Hoare triple.
\end{proposition}

\subsection[Leveled Houdini]{Leveled Houdini}
\label{sec:leveled-houdini}

The \LH algorithm takes as input a single-loop Hoare triple $\tau$ of the form in Equation~\ref{eq:loop-triple}, a finite set of \emph{candidate} clauses $S$, and a leveled candidate $\Kscr = (K_0, \hdots, K_m)$. All verification conditions and inductive cores in this section are relative to $\tau$. Assuming $\Kscr$ is an inductive core, \LH constructs an inductive core $\Kscr' = (K'_0, \hdots, K_n)$ with $m \leq n$ and $K_j \subseteq K'_j$ for all $j \leq m$. The output core may contain additional levels. We treat levels beyond the greatest tuple index of a leveled candidate as empty.

Intuitively, \LH iteratively applies ordinary \Houdini at each level $j$, starting from level $0$, and applying the level-$j$ initialization and preservation checks until reaching a fixed point. Once \LH reaches a fixed point at a given level, that level is no longer changed during the run. For level-$0$, the checks are the usual per-clause checks. At a positive level, these checks come with the additional assumptions discussed previously, which are not provided by the standard \Houdini algorithm. Clauses in the candidate set $S$ which do not contain prophecy variables are first tried at level $0$, while those containing prophecy variables are first tried at level $1$. A clause which fails either level-$j$ initialization or preservation is tried again at level $j+1$, unless $j>0$ and the updated core contains no clauses at level $j$ or above. This is because the additional assumptions may cause a clause to pass its checks at a higher level despite failing at a lower one.

\input{figures/leveledhoudini.tex}

Pseudocode for \LH is provided in Algorithm~\ref{alg:leveled-houdini}. The outer loop iterates through levels in increasing order, with $k$ denoting the current level. We write $\ell(c)$ to denote the initial level of clause $c$ (either $0$ or $1$), as described above. The set $P$ contains all clauses from $S$ which do not already occur at some level in $\Kscr$. On the $k^{th}$ iteration of the outer loop, $R$ contains those clauses to be considered for addition to level $k$ of the core. The function $\Extend(\Kscr, k, R)$ returns the result of adding every clause in $R$ to the $k^{th}$ level of $\Kscr$. The algorithm assumes access to an oracle for \Whiel assertion entailments. As in classical Houdini, the body of the outer loop first removes all clauses failing initialization for the corresponding level, followed by an inner loop removing all clauses which fail preservation. Clauses which survive this pruning process are added to the core at the corresponding level.

\subsubsection*{Correctness}
Ordinary Houdini's pruning argument entails that each level also produces the largest subset of the clauses considered at that level which pass the initialization and preservation checks for that level. Furthermore, their inclusion at each level only adds to the hypotheses of the initialization and preservation checks for clauses already in the core. The inner loop always terminates because each iteration either removes a clause from $R$, or else the guard fails on the next iteration. The outer loop halts because any iteration for a level above $m$ either removes a clause from $P$, or the loop is broken on line 14. Furthermore, if $\term_{\Kscr'}$ holds for the returned core $\Kscr'$, then the input Hoare triple is valid by Proposition~\ref{prop:ladder-rule}.

%% file: figures/leveledhoudini.tex
\begin{algorithm}
\caption{$\LH(\tau,S,\Kscr)$}
\label{alg:leveled-houdini}
\begin{algorithmic}[1]
\Require Single-loop Hoare triple $\tau$; finite candidate clause set $S$; Inductive core $\Kscr = (K_0, \hdots, K_m)$;
\Ensure An inductive core extending $\Kscr$
\State $P \gets S \setminus \bigcup\Kscr \seq k \gets 0$
\While{$P \neq \emptyset$}
    \State $R \gets \{ c \in P \mid \ell(c) \leq k \} \seq \Rold \gets \emptyset$
    \For{$c \in R$}
        \State \textbf{if} $\init^k_{\Kscr,c}$ is invalid \textbf{then} $R \gets R \setminus \{c\}$
    \EndFor
    \While{$R \neq \Rold$}
        \State $\Rold \gets R$
        \State $\Ascr \gets \Extend(\Kscr, k, R)$
        \For{$c \in R$}
            \If{$\step^k_{\Ascr,c}$ is invalid}
                \State $R \gets R \setminus \{ c \}$
                \State \textbf{break}
            \EndIf
        \EndFor
    \EndWhile
    \State $\Kscr \gets \Extend(\Kscr,k,R) \seq P \gets P \setminus R$
    \State \textbf{if} $k > 0$ and $K_j=\emptyset$ for every $j\geq k$ \textbf{then break}
    \State $k\gets k+1$
\EndWhile
\State \Return $\Kscr$
\end{algorithmic}
\end{algorithm}

%% file: sections/s5-system.tex
\section{Implementation Overview}
\label{sec:implementation}

The \Whiel system implements the pipeline described in Figure~\ref{fig:system-overview}. It takes as input a Hoare triple, conducts a search for a leveled invariant or a counterexample input state, and, when the search succeeds, certifies the solution in \Lean. \Lean certification serves as a guarantee of soundness; consequently, Theorem~\ref{thm:whv-coRE-complete} implies that \Whiel is incomplete. To mitigate the risk of non-termination, the implementation incorporates a number of resource limits. These include configurable time limits bounding the overall search time, the time allotted to \Vampire solver calls, the execution of candidate counterexamples, and the construction of \Lean certificates. Exhausting the overall search time leaves the triple unresolved, while exhausting the certification time leaves the verifier's answer uncertified. A Rust-based controller manages the interactions between the \Lean library, \Vampire, and a Python harness for the LLM-based proposer. We describe the components below, with the experimental parameters deferred until Section~\ref{sec:experiments}.

\subsection{\Lean Foundation and Preprocessing}
\label{sec:lean-foundation}
\Whiel's \Lean library formalizes database instances, syntax and semantics for \RA, \Whiel assertions and commands, and soundness of the Hoare logic principles introduced in Section~\ref{sec:verification}. It also implements several operations used by the synthesis algorithm, including $\wp$ computation, preprocessing, and various syntactic operations on formulas such as prophecy variable substitution.

The preprocessor uses classical methods~\cite{harel1980folk}, which may introduce additional nullary relation symbols to the schema, to transform an arbitrary Hoare triple into \emph{single-loop form}, that is, the form of Equation~\ref{eq:loop-triple} with $B$ loop-free. A \Lean theorem proves that the resulting triple is \emph{equivalid}\footnote{Two Hoare triples are equivalid if either both are valid or both are invalid.} with the original triple. The search algorithm seeks an invariant over the preprocessed triple, while the equivalidity theorem transfers the result back to the original.

\subsection{Neurosymbolic Search with \AgentHoudini}
\label{sec:search}

\Whiel's search procedure is an interactive exchange between a \emph{proposer} and a \emph{verifier}. The verifier initially pushes the execution schema, the input Hoare triple, the prophecy schema, and the preprocessed Hoare triple to the proposer. On each round, the proposer has read-only access to tools provided by the verifier, including an assertion syntax well-formedness checker, a tool for inspecting a runtime record of verification condition check results and countermodels (discussed in more detail in Section~\ref{sec:checking-entailments}), and a tool to evaluate assertions on concrete instances. For our implemented \AgentHoudini proposer, these tools are optional, and we evaluated the effect in Section~\ref{sec:experiments}. The proposer then makes a $\Submission$, which is either a finite set of clauses over the prophecy schema (a \emph{candidate clause set}), or a concrete database instance (a \emph{candidate counterexample}). Note that we use \emph{countermodel} to refer to an instance which proves invalidity of a \Whiel entailment, while we use \emph{counterexample} to refer to an instance which proves invalidity of a Hoare triple.

An inductive core, initially empty, is maintained by the verifier and grows monotonically: once a clause is added to one of its levels, it is never removed during the remainder of the search. When a candidate clause set is submitted, the verifier runs an implementation of the \LH algorithm to expand the inductive core (see Section~\ref{sec:leveled-houdini}). If this core passes the leveled candidate termination check, then the search ends and we proceed to certification. Otherwise, the updated inductive core and the set of clauses from the previous proposal not accepted to the core are sent back to the proposer on the next round. A proposed clause $c$ not admitted to the core by \LH but to be reconsidered later is \emph{pending}; such clauses will be reconsidered at level $\ell(c)$ on the next round in which a candidate set is submitted. The proposer has the option to request that pending clauses be dropped (i.e., not reconsidered).

\input{figures/search.tex}

When a candidate counterexample is submitted, the verifier first checks that the instance satisfies the (original) precondition. If it does, then the verifier uses a \Lean-implemented \Whiel command interpreter to run the (original) command on the concrete state; if this halts, it checks if the output falsifies the (original) postcondition.

Pseudocode for our search procedure is provided in Algorithm~\ref{alg:search}, which runs after preprocessing and before certification\footnote{The pseudocode omits solver retry policies and other optimization heuristics.}. The special case of this abstract search procedure in which proposals are given by an LLM agent is the \AgentHoudini algorithm.

\subsection{Checking First-Order Validity with \Vampire}
\label{sec:checking-entailments}

\Whiel uses \Vampire in the place of the finite validity oracle assumed by \LH. This imposes some unavoidable limitations on the system. First, \Vampire checks validity of first-order entailments over \emph{arbitrary} structures, whereas \Whiel entailments correspond to first-order entailments over \emph{finite} structures. A valid \FOL entailment is also finitely valid, but the reverse does not hold. Thus \Vampire is \emph{sound} but \emph{incomplete} for \Whiel entailments.

For each verification condition checked, \Lean code translates the \Whiel entailment into an equivalent\footnote{Modulo details involving empty structures, which are handled in the implementation.} first-order entailment over finite structures. This \FOL entailment is then passed to \Vampire by the Rust controller to execute a concurrent proof and finite countermodel search. Both the proof search and finite countermodel search are subject to the same \emph{solver time budget}. There are three possible outcomes: success (validity via proof), failure (invalidity via finite countermodel), or inconclusive. The latter outcome occurs when the solver time budget is exhausted or when \Vampire encounters an error, such as running out of memory. When pruning clauses in \LH, our implementation treats inconclusive checks as failures. As a result, this also limits the theoretical guarantee that the core contains the largest inductive subset of clauses considered at each level. Similarly, inconclusive termination checks may also occur even when the core is actually sufficient for validity.

\subsection{Solution Certification and Trusted Codebase}
\label{sec:certification-trust}

Certification occurs \emph{after} the search procedure has returned either a leveled invariant or a validated counterexample. For the former, the certifier constructs \Lean files containing formalizations of the verification conditions. It then invokes \Vampire's \LeanCheck, which produces a proof that is reconstructed in \Lean using the \VampLean tool\footnote{Our current system applies some small proof transformations to the generated proofs in order to improve compilation time and replace invocations of the \texttt{bv\_decide} SAT checker with kernel-checked LRAT replay to avoid dependence on axioms introduced by native evaluation.}. These certified proof reconstructions are combined with theorems from the \Whiel \Lean library's Hoare logic formalization and preprocessing correctness theorem to produce a validity proof for the input Hoare triple. Similarly, counterexamples accepted by the verifier have their validation re-run in the \Lean kernel, producing a proof of invalidity for the input Hoare triple. The emitted theorems in these \Lean \emph{certificates} are named \texttt{input\_hoare\_triple\_valid} or {input\_hoare\_triple\_invalid}. Certification does not rely on an LLM agent, and solutions can be certified independently of the search process that produced them. The certifier rejects any dependencies of the emitted theorems on axioms other than \Lean's standard axioms. A failure of proof reconstruction or counterexample replay during certification produces no certificate.

The purpose of the certifier is to produce a minimal \emph{trusted code base}. A successfully certified solution for the \Whiel Hoare validity problem does not require trust in the search algorithm, Rust controller, proposer, \Vampire, \VampLean, or even the code that generates the certificate. Instead, the only requirement for confidence in a certified solution is trust in the \Lean kernel and trust in the fidelity of the \Lean library's formalization of \Whiel commands and the definition of Hoare validity. To this end, we provide a file \texttt{AUDIT.md} in the \Whiel repository with a map for auditing the \Lean formalization.

%% file: figures/search.tex
\begin{algorithm}
\caption{\Whiel's Search Algorithm}
\label{alg:search}
\begin{algorithmic}[1]
\Require Original triple $\tau$; preprocessed triple $\tau'$; search time budget
\Ensure Accepted core, validated counterexample, or unresolved
\State $\PushTask(\tau,\tau')$
\State $\Kscr \gets ( \emptyset ) \seq \Pending \gets \emptyset$
\While{search time budget remains}
    \State $\Submission \gets \Propose(\Kscr, \Pending)$
    \If{$\Submission$ is a candidate clause set $S$}
        \State $\Kscr \gets \LH(\tau', S \cup \Pending, \Kscr)$
        \State $\Pending \gets (S \cup \Pending) \setminus \bigcup\Kscr$
        \State \textbf{If} $\term_{\Kscr}=\valid$ \textbf{then} \Return $\Kscr$
    \ElsIf{$\Submission$ is a candidate counterexample $s$}
        \State \textbf{If} $\Validate(\tau,s)$ succeeds \textbf{then} \Return $s$
    \EndIf
\EndWhile
\State \Return unresolved
\end{algorithmic}
\end{algorithm}

%% file: sections/s6-experiments.tex
\providecommand{\toolname}[1]{\textsf{#1}}

\section{Experiments}
\label{sec:experiments}

We evaluate \Whiel on a benchmark of 86 verification tasks that we curated (Section~\ref{sec:exp-benchmark}) and ask the following questions:
(1) {\bf RQ1 (Comparison of \Whiel with Baselines):} Does our method \Whiel solve tasks that the same LLM cannot solve by writing \Lean proofs directly?
(2) {\bf RQ2 (Effect of Skills and Tools):} Do curated skills and MCP tools help the proposer solve more tasks, or solve them in fewer rounds, in less time, or at a lower cost?
(3) {\bf RQ3 (Impact of Stronger Models in the Proposer):} When stronger models are used as a proposer, does \Whiel solve more tasks, and which tasks need them?
(4) {\bf RQ4 (Effect of Prophecy Variables):} How do tasks requiring prophecy variables compare with those that do not need them?
(5) {\bf RQ5 (Certification by \Whiel):} What is the additional overhead for using the optional certifier provided by \Whiel?

\subsection{Benchmark}
\label{sec:exp-benchmark}
To our knowledge, no existing benchmark targets the verification of imperative relational programs against relational specifications.
Therefore, we construct a benchmark of 86 verification tasks.
Those tasks are drawn from classical database problems (e.g., \Datalog programs~\cite{abiteboul1995foundations}, query reformulation using views~\cite{duschka1997answering}, DRed~\cite{gupta1993maintaining} and the chase~\cite{maier1979testing,benedikt2017benchmarking}), real systems (e.g., PostgreSQL~\cite{postgresql}, NetworkX~\cite{hagberg2008networkx} and Souffl\'e~\cite{herbert2016souffle}), and LLM-generated programs.
Each task is a Hoare triple $\Hoare{P}{C}{Q}$ over a \Whiel program, together with its ground-truth verdict: valid or invalid.
The suite contains 69 valid and 17 invalid tasks and instantiates the problem classes of Section~\ref{sec:applications}, including recursive query equivalence and containment, evaluation strategies, incremental view maintenance, and relational correctness properties.
A detailed description of the benchmark and 86 tasks is available in Appendix~\ref{app:benchmark} and the benchmark is available in the \Whiel git repository \cite{whielgit}.

Table~\ref{tab:benchmark-kinds} groups the tasks by the type of their postcondition.
Most state query equivalence (64 tasks), e.g., the naive and semi-naive evaluation of a \Datalog program, query rewriting with views, or query containment (18 tasks).
The remaining four state a property of a single computed relation, such as transitivity (correctness, 2 tasks), or combine claims through an implication or a disjunction (2 tasks).
Table~\ref{tab:benchmark-structure} summarizes the size of the tasks; 41 of the 69 reference invariants use prophecy relations, 34 of them in equivalence tasks.

Every task is backed by a \Lean certificate:
for a valid task, the verification conditions of a reference invariant are proved in \Lean; for an invalid task, the \Lean kernel replays a counterexample. In our experiments, the proposer never sees the reference answers; we use them only as ground truth, to confirm the verdicts that \Whiel returns and to characterize the difficulty of the tasks.

\begin{table}[t]
  \centering
  \caption{The benchmark summary statistics.}
  \vspace{-1em}
  \label{tab:benchmark}
  \footnotesize
  \setlength{\tabcolsep}{2.5pt}
  \begin{subtable}[t]{0.49\columnwidth}
    \centering
    \caption{Tasks by postcondition type.}
    \label{tab:benchmark-kinds}
    \begin{tabular}{lccc}
      \toprule
      Type & Tasks & Valid & Invalid \\
      \midrule
      Equivalence & 64 & 52 & 12 \\
      Containment & 18 & 14 & 4 \\
      Correctness & 2 & 1 & 1 \\
      Implication & 2 & 2 & 0 \\
      \midrule
      Total & 86 & 69 & 17 \\
      \bottomrule
    \end{tabular}
  \end{subtable}\hfill
  \begin{subtable}[t]{0.49\columnwidth}
    \centering
    \caption{Structure of the tasks.}
    \label{tab:benchmark-structure}
    \setlength{\tabcolsep}{1.8pt}
    \begin{tabular}{lccc}
      \toprule
      & Min & Median & Max \\
      \midrule
      \# While loops & 1 & 2 & 6 \\
      \# Relation symbols & 3 & 9 & 32 \\
      \# Assignments & 4 & 12 & 43 \\
      \# Invariant clauses$^{*}$ & 2 & 10 & 33 \\
      \bottomrule
    \end{tabular}
  \end{subtable}

  \smallskip
  \raggedright\footnotesize $^{*}$Over the reference invariants of the 69 valid tasks.
  \vspace{-2mm}
\end{table}

\subsection{Experimental Setup}\label{subsec:exp-setup}
Experiments ran on a Linux server with an AMD EPYC 7R13 processor (48 cores, 2.6\,GHz) and 256\,GB RAM. 
\Whiel uses \Lean~4 and \Vampire\footnote{The \Whiel git repository~\cite{whielgit} specifies the exact builds of both tools.}.
The proposer is an OpenAI Codex CLI agent~\cite{codexcli} that calls remote model APIs from a sandbox hiding benchmark solutions and earlier results.
Four tasks run in parallel, each with at most 12 single-threaded solver processes (8 \Vampire and 4 \Lean).

\noindent
\textbf{Methods.}
We evaluate \textbf{\Whiel}, whose search is with \AgentHoudini (Section~\ref{sec:search}): an LLM agent proposes and the verifier checks. 
In the default configuration, which we call simply \Whiel{} throughout this section, the agent can only submit invariant clauses or a counterexample and receive the verifier's feedback.
We also built two extensions of this interface with {\em MCP tools} and {\em skills}, designed to help the agent exploit the verifier's feedback and domain knowledge (see Appendix~\ref{app:tools-skills}). Each extension can be enabled independently, which gives four configurations: \textbf{\Whiel}, \textbf{\WhielSkills}, \textbf{\WhielTools} and \textbf{\WhielToolsSkills}.
(1) \textbf{Tools} are six read-only tools (Section~\ref{sec:search}) that the verifier exposes to the agent through the {\em Model Context Protocol (MCP)}~\cite{mcp}.
(2) \textbf{Skills} are six curated tutorials on constructing and repairing invariants, which the agent reads on demand.
Each is a written procedure: when it applies, what to inspect in the task, how to construct candidate clauses, and how to read the next failure.
For example, \toolname{Ladder strategy} explains how to build a leveled invariant bottom-up, from facts that hold at every iteration to clauses that need facts about the exit state.
In the configurations that provide them the prompt lists the available tools and skills with one-line descriptions, and the agent decides whether to use them.

\noindent
\textbf{Baselines.}
To our knowledge, no existing tool automatically verifies imperative relational programs with while loops, so we compare \Whiel against two baselines that use the same LLM (i.e., GPT-5.5).
(i)~\emph{\directlean}: 
an LLM agent receives the task as a \Lean Hoare-validity statement with the semantics, proof lemmas and proof interface, and may read the library sources, compile \Lean code and revise its proof in an isolated sandbox, without access to \AgentHoudini, the \Whiel verifier or the reference answers. It submits a verdict with a complete \Lean proof of the statement or of its negation, which counts only if the \Lean kernel checks it in full\footnote{That is, with no unproved placeholders (\texttt{sorry}) or extra axioms.}.
(ii) \emph{\whieloneround}: only the first round of each \Whiel run, one proposal followed by invariant or counterexample checking, without verifier feedback.

\noindent
\textbf{LLMs and budgets.}
Unless stated otherwise, the agent uses GPT-5.5~\cite{gpt55}, to show the effectiveness of \Whiel with an LLM at moderate cost.
In the model-scaling experiment (Section~\ref{sec:rq3}), the tasks that GPT-5.5 leaves unsolved are passed to stronger models, GPT-5.6 Sol~\cite{gpt56sol} and then GPT-6 Astra~\cite{gpt6astra}; each model attempts only the tasks left unsolved by the previous one, from scratch, so the coverage we report is cumulative. We set the reasoning effort to \texttt{medium} for all models.
Each attempt has a 600\,s search budget for the proposer's turns and all verifier checks and ends when the verifier accepts an answer (invariant or  counterexample) or when the budget runs out; 
\directlean runs without a time limit. %

\noindent
\textbf{Metrics.}
We report three metrics.
(i) \emph{Solved tasks}: the number and fraction of the 86 tasks that a method solves.
\Whiel solves a task when its verifier accepts the answer (an invariant for validity or a counterexample for invalidity); \directlean solves a task when \Lean accepts its proof.
(ii) \emph{Search time}: the elapsed time per task, which covers the agent's turns and all checks but not certification.
(iii) \emph{Rounds}: the number of rounds per task, where a round is one submission from the proposer and its verification.

\subsection{RQ1: Comparison of \Whiel with Baselines
}
\label{sec:rq1}

\begin{table}[t]
\centering
\captionsetup{skip=4pt}
\caption{Comparison of \Whiel with baselines (using GPT-5.5). Time, rounds, and cost are averaged over all 86 tasks.}
\label{tab:main}
\begingroup
\small
\setlength{\tabcolsep}{1.8pt}
\renewcommand{\arraystretch}{1.1}
\begin{tabular}{lrrrrrr}
\hline
 & \multicolumn{3}{c}{\# Solved (\%)} & \multicolumn{3}{c}{Per-task averages} \\
\cline{2-7}
Method & Valid (69) & Invalid (17) & Total (86) & Time (s) & Rounds & Cost (\$) \\
\hline
\rowcolor{black!12}
\textbf{\Whiel} & \textbf{57}\,(83\%) & \textbf{16}\,(94\%) & \textbf{73}\,(85\%) & 189 & 2.95 & 0.49 \\
\hline
\begin{tabular}[c]{@{}l@{}}\Whiel-\\\textsc{OneRound}\end{tabular} & 32\,(46\%) & 14\,(82\%) & 46\,(53\%) & 58 & 1 & 0.14 \\
\hline
\begin{tabular}[c]{@{}l@{}}\directlean\end{tabular} & 4\,(6\%) & 16\,(94\%) & 20\,(23\%) & 398 & -- & 2.08 \\
\hline
\end{tabular}
\vspace{-2mm}
\endgroup

\end{table}

\Whiel outperforms both baselines in effectiveness. As Table~\ref{tab:main} shows, \directlean solves 20 tasks (23\%) even without a time limit. \Whiel solves 73 tasks (85\%) in less than half the time (189\,s vs.\ 398\,s) and at about a quarter of the cost (\$0.49 vs.\ \$2.08), and each of its verdicts rests on verification conditions proved by \Vampire or on a counterexample validated by the \Lean-implemented interpreter.
\directlean names the correct verdict for all 86 tasks. What it fails to produce is evidence for valid claims: it solves only 4 of the 69 valid tasks.
Of its 66 unsolved tasks, 45 do not compile, 19 compile but fail the axiom audit, and 2 submit no proof.\footnote{Of the 19, five leave \texttt{sorry} placeholders, six declare the goal as a new axiom, five prove only the tautology $\mathit{goal} \lor \neg\mathit{goal}$, and three omit the answer theorem. \texttt{sorry} is a \Lean placeholder that closes any proof goal without proving it.}

\whieloneround shows that even with one round, our framework does better than \directlean.
The model submits a single proposal, taking 15\% of the time of \directlean (58\,s vs.\ 398\,s) and 7\% of its cost (\$0.14 vs.\ \$2.08), and it only has to propose candidate clauses or a concrete counterexample: \LH keeps the inductive clauses, \Vampire discharges the verification conditions, and the \Lean-implemented interpreter validates counterexamples.
This division of labor raises the solve rate from 23\% to 53\%.
Iterating with verifier feedback adds 27 solved tasks, taking 3.1 rounds on average. In \Whiel with iterations and feedback, although the time increases, the success rate significantly increases even when skills and tools are not used.

\subsection{RQ2: Effect of Skills and Tools}
\label{sec:rq2}

To evaluate the effect of the two extensions of \Whiel with tools and/or skills, we compare four configurations using GPT-5.5.
As Figure~\ref{fig:rq2-acceptance} shows, \WhielSkills solves the most tasks over the full run, 77 of the 86 (90\%), while \Whiel, \WhielTools and \WhielToolsSkills have similar performance within the same search time, with 73~(85\%), 74~(86\%) and 73~(85\%) solved tasks. Of all tasks, 69 (80\%) are solved by all four configurations and 7 (8\%) by none of them.
Figure~\ref{fig:rq2-resources} shows that the four configurations behave alike on the tasks they solve: 1.5--1.8 rounds, 105--137\,s and 44--89 clause checks per task.
Despite similar overall coverage, the configurations differ in how quickly they find answers.
\WhielSkills solves 54 tasks in the first round, compared with 46--48 for the other configurations, and 66 by the second round, compared with 60 for \WhielTools and 53 for \Whiel.

As Figure~\ref{fig:rq2-resources} shows, compared with \Whiel, all extensions reduce the avg. number of rounds and avg. clause checks. Compared to \WhielTools, \WhielSkills is more effective as it reduces both avg. search time and avg. cost over \Whiel, while \WhielTools increases them.  In \WhielSkills, the model picks the skills that fit the task: it reads \emph{delta coupling} in 75\% of its model requests on tasks with semi-naive evaluation and never on other tasks, and three of the five tasks that it solves and \Whiel does not are among them.

\input{figures/rq2-acceptance}

\input{figures/rq2-resources}

\WhielToolsSkills
does not perform better than 
\WhielSkills and increases avg. cost from \$0.46 to \$0.80. It lowers the solved tasks from 77 to 73 and increases the avg. rounds from 2.09 to 2.35.
We see two likely reasons. First, these extensions overlap: both steer the model toward the missing clauses, skills by describing invariant patterns up front and tools by exposing concrete countermodels, so the second adds little guidance to the first. Second, their overheads add up. With both, the model makes 7.7 tool and skill calls per model request instead of 3.5 and processes 175k tokens per request instead of 79k; its largest context per request averages 34.5k tokens instead of 21.2k as tool replies accumulate. Each round then takes 84\,s instead of 76\,s. The longer context may also dilute useful hints. 

\subsection{RQ3: Impact of Stronger Models in Proposer}
\label{sec:rq3}

\input{figures/rq3-escalation}

As shown in Figure~\ref{fig:rq3-escalation}, we
run every task that the GPT-5.5 model leaves unsolved, first  with  GPT-5.6 Sol, then with GPT-6 Astra. The coverage, i.e., the number of solved tasks, 
increases from 73 to 81 (GPT-5.6 Sol), and to 85 of the
86 tasks for \Whiel, and the other three configurations also reach 85 with Astra.
The tasks that need a stronger model are the same at the hard end in every configuration: four tasks resist GPT-5.5 and Sol everywhere, and Astra solves three of them, namely red/blue alternating paths (linear vs.\ nonlinear), DRed deletion vs.\ recomputation, and Souffl\'e's topological ordering (naive vs.\ semi-naive).
The last one is invalid: GPT-5.5 and Sol run out of time, whereas Astra submits a counterexample in its first model request in all four configurations.

\subsection{RQ4:
Effect of Prophecy Variables
}
\label{sec:rq4}

One obvious source of difficulty in the tasks is {\bf whether a prophecy variable is necessary} (Section~\ref{sec:prophecy}); the success rates for different models are shown below:

{
    \centering
\small

    \begin{tabular}{|c|c|c|c|c|}
    \hline
        & GPT 5.5 & GPT 5.6 & GPT 6  & Unsolved\\\hline
        Prophecy (total 41) &  30  (73\%) & 7  (17\%) & 3 (7\%) & 1 (3\%) \\
        No prophecy (total 28) &  27  (96\%) & 1  (4\%) & - & - \\\hline
    \end{tabular}
}

As expected, if prophecy variables are not needed, the success rate is high even for GPT 5.5 and the overall success rate is 100\%.
For some tasks requiring prophecy variables, a clause over terminal values cannot be read off either program: the proposer has to guess which facts about the terminal state hold.
The hardest tasks nest such facts.
Both tasks whose every known invariant has clauses at level~2 (Section~\ref{sec:leveled-invariants}) need a stronger model. For example, GPT 5.6 Sol solves \emph{Red/blue closure vs.\ \Datalog path views}, and no model solves \emph{Warshall sweep vs.\ PostgreSQL recursive union} (see tasks 4040 and 5034 in the git repository~\cite{whielgit}) within the budget, although GPT 6 Astra proposes its nested facts in its first round.
There, \Vampire seems to be the bottleneck: even given the reference invariant, the verifier needs 333\,s of the 600\,s budget to prove it.

\subsection{RQ5: Certification by \Whiel}
\label{sec:exp-certification}
Running the certifier is an optional process in \Whiel, depending on the need of the application. After all search experiments were complete, we ran the certifier on the 340 verifier-accepted answers from all four configurations utilized during search (85 per configuration). We utilized a 300\,s overall time limit, a 300\,s internal \Vampire call allowance, and memory limits of 10\,GB per process. Each certification attempt used 12 physical cores and 12 solver workers, with two attempts running concurrently on disjoint cores. This certified 293/340 answers (86.2\%): 225/272 validity answers and all 68 invalidity answers. Among successful attempts, the certification time had mean 72.2\,s, median 53.1\,s, and maximum 295.8\,s; sampled max single-process memory usage had mean 2.753\,GB, median 2.540\,GB, and maximum 7.591\,GB. The remaining attempts resulted in 41 timeouts and six memory-limit failures. These resource failures of the certifier reflect known difficulties for \Lean with elaborating and type-checking large \VampLean-generated proofs~\cite{bodingbauer2026vamplean}; improving resource usage here is a target for future work.

%% file: figures/rq2-acceptance.tex
\begin{figure}[t]
  \centering
  \includegraphics[width=\columnwidth]{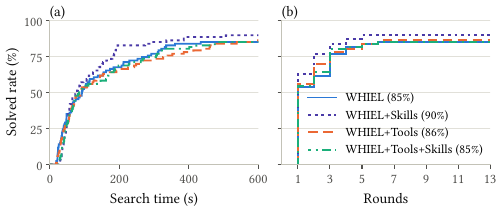}
  \caption{Solved rate of the four configurations of \Whiel, without and with (+ tools and/or skills) on 86 tasks; GPT-5.5.}
  \label{fig:rq2-acceptance}
\end{figure}%

%% file: figures/rq2-resources.tex
\begin{figure}[t]
  \centering
  \includegraphics[width=\columnwidth]{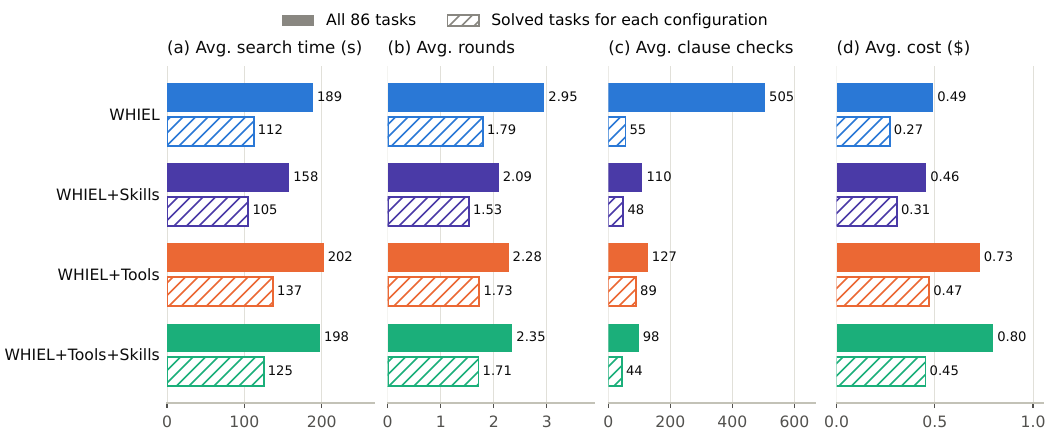}
  \caption{
  Average search time, rounds, clause checks and cost per task for the four  \Whiel configurations over all 86 tasks (solid) and solved tasks
  (hatched); GPT-5.5. Avg. search time in (a) for all tasks includes 600 s for timed-out tasks.
  }
  \label{fig:rq2-resources}
\end{figure}%

%% file: figures/rq3-escalation.tex
\begin{figure}[t]
  \centering
  \includegraphics[width=\columnwidth]{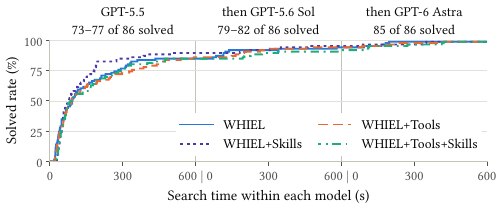}
  \vspace{-2em}
  \caption{Solved rate and search time of the four \Whiel configurations;
  the tasks left unsolved by
  one model are passed to the next: GPT-5.5, then GPT-5.6 Sol, then GPT-6 Astra.
  }
  \label{fig:rq3-escalation}
\end{figure}%

%% file: sections/s7-conclusion.tex
\section{Conclusion and Future Work}
\label{sec:conclusion}

We have presented the \Whiel system, which combines LLM-guided invariant and counterexample synthesis with a symbolic verification pipeline employing a novel \LH algorithm and an independent certification procedure. Our experiments demonstrate the value of this neurosymbolic architecture on a curated benchmark of examples motivated by hard database problems. Future work includes developing a more comprehensive relational programming benchmark, refining the agent-based synthesis pipeline, and reducing the time and memory requirements of the certifier. We also plan to study extensions of the language with fresh-value creation, arithmetic, aggregation, and semiring-annotated data.

%% file: sections/appendix.tex
\section{Additional Experimental Details}

\begin{table*}[!htbp]
  \centering
  \captionsetup{skip=4pt}
  \caption{The 86 tasks by source, with their identifiers in the \Whiel git repository~\cite{whielgit}.
  Claim: equivalence (Eq), containment (Co), correctness (Cr) or implication (Im). V/I: valid or invalid. P: the reference invariant uses prophecy variables. E: the task verifies an evaluation strategy (Appendix~\ref{app:rq4}). 
  lfp/gfp: least/greatest fixpoint of the named rules. The Example numbers are arbitrary. 
  The corresponding file in git has a prefix of `Example', e.g., `Example0001'.}
  \label{tab:app-tasks}
  \footnotesize
  \setlength{\tabcolsep}{3pt}
    \begin{tabular}[t]{@{}l>{\raggedright\arraybackslash}p{0.315\textwidth}cccc@{}}
    \toprule
    ID & Task & Claim & V/I & P & E \\
    \midrule
    \multicolumn{6}{@{}l}{\emph{Classic \Datalog programs (21)}} \\
    0001 & Naive TC is transitively closed & Cr & V & $\bullet$ &  \\
    0013 & Naive TC under the false bound $T \subseteq E$ & Cr & I &  &  \\
    0106 & Left-linear TC vs.\ lfp of nonlinear TC & Eq & V & $\bullet$ &  \\
    0107 & Nonlinear TC vs.\ lfp of left-linear TC & Eq & V &  &  \\
    0108 & Right-linear TC vs.\ lfp of left-linear TC & Eq & V &  &  \\
    0109 & Linear red/blue alternating paths $\subseteq$ TC & Co & V &  &  \\
    0115 & All paths via linear odd/even vs.\ lfp of TC & Eq & V &  &  \\
    0117 & All paths via nonlinear odd/even vs.\ lfp of TC & Eq & V &  &  \\
    0119 & All paths via linear odd/even vs.\ lfp of nonlinear TC & Eq & V & $\bullet$ &  \\
    0123 & Nonlinear odd/even vs.\ lfp of linear odd/even & Eq & V &  &  \\
    0125 & Linear odd/even vs.\ lfp of nonlinear odd/even & Eq & V & $\bullet$ &  \\
    0128 & Nonlinear red/blue alternating paths $\subseteq$ TC & Co & V &  &  \\
    0130 & Path-system depth parities $\subseteq$ lfp of the rules & Co & V &  &  \\
    0132 & Odd/even paths: linear vs.\ nonlinear & Eq & V & $\bullet$ &  \\
    0133 & All paths via odd/even paths: linear vs.\ nonlinear & Eq & V & $\bullet$ &  \\
    0134 & Red/blue alternating paths: linear vs.\ nonlinear & Eq & V & $\bullet$ &  \\
    0136 & Right-linear TC vs.\ nonlinear odd/even paths & Eq & V & $\bullet$ &  \\
    0137 & Nonlinear TC vs.\ linear odd/even paths & Eq & V & $\bullet$ &  \\
    0138 & Nonlinear TC vs.\ nonlinear odd/even paths & Eq & V & $\bullet$ &  \\
    0162 & Reachability game vs.\ lfp of its rules & Eq & V &  &  \\
    0163 & Safe set by shrinking iteration vs.\ gfp & Eq & V &  &  \\
    \midrule
    \multicolumn{6}{@{}l}{\emph{Query reformulation using views (10)}} \\
    4001 & Directed triangles vs.\ path-view reformulation & Eq & V &  &  \\
    4002 & TC vs.\ length-two path-view reformulation & Eq & I &  &  \\
    4003 & Path-view reformulation $\subseteq$ TC & Co & V & $\bullet$ &  \\
    4004 & Path-view reformulation vs.\ even-length paths & Eq & V & $\bullet$ &  \\
    4034 & Reformulation over sound odd/even views $\subseteq$ TC & Im & V & $\bullet$ &  \\
    4035 & Good-interior paths vs.\ edge-view reformulation & Eq & V & $\bullet$ &  \\
    4036 & Airline reachability vs.\ three views ($\subseteq$) & Co & V & $\bullet$ &  \\
    4037 & Airline reachability vs.\ three views ($\supseteq$) & Co & I &  &  \\
    4040 & Red/blue closure vs.\ \Datalog path views & Eq & V & $\bullet$ &  \\
    4041 & TC vs.\ reformulation over exact odd/even views & Eq & V & $\bullet$ &  \\
    \midrule
    \multicolumn{6}{@{}l}{\emph{Specification vs.\ procedure (11)}} \\
    5014 & SPARQL property paths vs.\ \Datalog closure & Eq & V &  & $\bullet$ \\
    5015 & SPARQL $p^{*}$ vs.\ closure plus identity on endpoints & Eq & I &  & $\bullet$ \\
    5016 & DRed deletion vs.\ recomputation & Eq & V & $\bullet$ & $\bullet$ \\
    5017 & DRed without rederivation vs.\ recomputation & Eq & I &  & $\bullet$ \\
    5018 & Chase: one dependency at a time vs.\ parallel & Eq & V &  & $\bullet$ \\
    5019 & Chase: one pass vs.\ parallel & Eq & I &  & $\bullet$ \\
    5020 & RDFS: $\rho$df rules vs.\ full rule set & Eq & V &  &  \\
    5021 & RDFS: $\rho$df vs.\ full, reflexive triples dropped & Eq & I &  &  \\
    5022 & SQL recursion limit: completion implies closure & Im & V & $\bullet$ &  \\
    5023 & Win-move: alternating fixpoint vs.\ attractor & Eq & V & $\bullet$ & $\bullet$ \\
    5024 & Win-move: odd limit claimed as the win set & Eq & I &  & $\bullet$ \\
    \bottomrule
  \end{tabular}\hfill
    \begin{tabular}[t]{@{}l>{\raggedright\arraybackslash}p{0.315\textwidth}cccc@{}}
    \toprule
    ID & Task & Claim & V/I & P & E \\
    \midrule
    \multicolumn{6}{@{}l}{\emph{Production programs (7)}} \\
    2004 & Souffl\'e magic sets vs.\ original program & Eq & V & $\bullet$ &  \\
    2005 & npm: restricted- $\subseteq$ full-edge reachability & Co & V &  &  \\
    2006 & Cilium/Batfish: filtered $\subseteq$ raw connectivity & Co & V &  &  \\
    2007 & NuSMV: $E[f\,U\,g] \subseteq EF\,g$ & Co & V &  &  \\
    2010 & Semi-naive TC vs.\ lfp of TC & Eq & V &  & $\bullet$ \\
    2011 & Incremental reachability maintenance vs.\ lfp & Eq & V &  & $\bullet$ \\
    2012 & Go GC frontier marking vs.\ lfp of marking & Eq & V &  & $\bullet$ \\
    \midrule
    \multicolumn{6}{@{}l}{\emph{Algorithm pairs (22)}} \\
    5001 & PostgreSQL recursive union vs.\ naive closure & Eq & V & $\bullet$ & $\bullet$ \\
    5002 & Datafrog semi-naive vs.\ naive evaluation & Eq & V & $\bullet$ & $\bullet$ \\
    5003 & Worklist BFS vs.\ naive reachability & Eq & V & $\bullet$ & $\bullet$ \\
    5004 & Casbin role closure vs.\ recursive has-role query & Eq & V & $\bullet$ & $\bullet$ \\
    5005 & Incremental reachability view vs.\ recomputation & Eq & V & $\bullet$ & $\bullet$ \\
    5006 & Semi-naive TC: appended vs.\ prepended delta & Eq & V & $\bullet$ & $\bullet$ \\
    5007 & TC by repeated squaring vs.\ linear closure & Eq & V & $\bullet$ & $\bullet$ \\
    5012 & Casbin depth-bounded roles $\subseteq$ unbounded closure & Co & V & $\bullet$ &  \\
    5013 & Casbin depth-bounded roles vs.\ unbounded closure & Eq & I &  &  \\
    5030 & NetworkX per-source TC vs.\ PostgreSQL & Eq & V & $\bullet$ & $\bullet$ \\
    5031 & NetworkX per-source TC, cutoff 2, vs.\ PostgreSQL & Eq & I &  & $\bullet$ \\
    5032 & Incremental TC under insertions vs.\ PostgreSQL & Eq & V & $\bullet$ & $\bullet$ \\
    5033 & Souffl\'e magic sets vs.\ PostgreSQL & Eq & V & $\bullet$ & $\bullet$ \\
    5034 & Warshall sweep vs.\ PostgreSQL recursive union & Eq & V & $\bullet$ & $\bullet$ \\
    5035 & Warshall sweep over two vertices vs.\ PostgreSQL & Eq & I &  & $\bullet$ \\
    5036 & NetworkX DAG TC vs.\ PostgreSQL & Eq & V & $\bullet$ & $\bullet$ \\
    5037 & Go GC frontier marking vs.\ Souffl\'e closure & Eq & V & $\bullet$ & $\bullet$ \\
    5038 & OpenFGA Check vs.\ ListObjects & Eq & V & $\bullet$ & $\bullet$ \\
    5039 & Exclusion per object $\subseteq$ exclusion on closure & Co & V &  &  \\
    5040 & Exclusion per object vs.\ exclusion on closure & Eq & I &  &  \\
    5041 & PostgreSQL FK cascade vs.\ declarative delete & Eq & V & $\bullet$ & $\bullet$ \\
    5042 & Over-approximating RESTRICT check vs.\ PostgreSQL & Eq & I &  & $\bullet$ \\
    \midrule
    \multicolumn{6}{@{}l}{\emph{Souffl\'e benchmark programs (4)}} \\
    5008 & Souffl\'e tc: naive vs.\ semi-naive & Eq & V & $\bullet$ & $\bullet$ \\
    5009 & Souffl\'e family: naive vs.\ semi-naive & Eq & V & $\bullet$ & $\bullet$ \\
    5010 & Souffl\'e orbits: naive vs.\ semi-naive & Eq & V & $\bullet$ & $\bullet$ \\
    5011 & Souffl\'e topological ordering: naive vs.\ semi-naive & Eq & I &  & $\bullet$ \\
    \midrule
    \multicolumn{6}{@{}l}{\emph{LLM-generated programs (11)}} \\
    1030 & Foreign-key consistent core of sharded rows vs.\ gfp & Eq & V &  &  \\
    1061 & Same generation vs.\ lfp of its rules & Eq & V &  &  \\
    1081 & Audited delegation $\subseteq$ authorization closure & Co & V & $\bullet$ &  \\
    1088 & Forward dependency closure vs.\ reverse resolution & Eq & V &  &  \\
    1090 & Taint reachability: ternary vs.\ projected frontier & Eq & V &  &  \\
    1091 & Delegation reachability $\subseteq$ closed policy view & Co & V &  &  \\
    1093 & Lineage triples $\subseteq$ closed ternary cover & Co & V &  &  \\
    1094 & Routing table $\subseteq$ closed distance view & Co & V &  &  \\
    3001 & Alternating audited delegation $\subseteq$ jointly certified & Co & I &  &  \\
    3009 & Certified delegation $\subseteq$ audit-bound authorization & Co & I &  &  \\
    3014 & Auditor-checked $\subseteq$ resource-checked elevation & Co & I &  &  \\
    \bottomrule
  \end{tabular}
\end{table*}

\subsection{\texorpdfstring{Benchmark (Section~\ref{sec:exp-benchmark})}{Benchmark}}
\label{app:benchmark}
In this section, we describe where the tasks of the benchmark come from and list all 86 tasks.
As in Section~\ref{sec:exp-benchmark}, the tasks come from classical database problems, real systems and LLM-generated programs; Table~\ref{tab:app-sources} splits the first two further by what the tasks compare, and Table~\ref{tab:app-tasks}
lists all 86 tasks, including those discussed in Section~\ref{sec:experiments}.
Among the classical problems, the \Datalog tasks compare different formulations of the same recursive query, such as linear and nonlinear transitive closure or odd/even paths, either by running both or by comparing one with the least fixpoint of the other's rules; a few check a single program against a property, e.g., that the computed closure is transitive.
The query-reformulation tasks compare a query with its reformulation over views~\cite{duschka1997answering}, and the specification-versus-procedure tasks compare a procedure from a standard or a paper, such as SPARQL property-path evaluation, DRed~\cite{gupta1993maintaining}, the chase~\cite{benedikt2017benchmarking} or the alternating fixpoint of the well-founded semantics, with a declarative program.
Among the real systems, the production tasks compare one program from the code or documentation of a system, such as Souffl\'e, PostgreSQL, npm, NuSMV or the Go runtime, with a declarative specification, usually the least fixpoint of \Datalog rules; the algorithm pairs compare two algorithms for the same query, e.g., NetworkX's or SciPy's transitive closure against PostgreSQL's recursive union; and the Souffl\'e tasks compare the naive and semi-naive evaluation of four programs from Souffl\'e's benchmark suite~\cite{herbert2016souffle}.
The code of these systems is translated into \Whiel programs.
Finally, 9 of the 11 LLM-generated tasks compare two programs whose inputs satisfy tuple-generating dependencies stated in the precondition.

\begin{table}[!htbp]
  \centering
  \captionsetup{skip=4pt}
  \caption{Benchmark tasks by source.
  }
  \label{tab:app-sources}
  \setlength{\tabcolsep}{3pt}
    \begin{tabular}{@{}lrrr@{}}
    \toprule
    Source & Tasks & Valid & Invalid \\
    \midrule
    \emph{Classical database problems} & 42 & 34 & 8 \\
    \quad Classic \Datalog programs & 21 & 20 & 1 \\
    \quad Query reformulation using views & 10 & 8 & 2 \\
    \quad Specification vs.\ procedure & 11 & 6 & 5 \\
    \emph{Real systems} & 33 & 27 & 6 \\
    \quad Production programs & 7 & 7 & 0 \\
    \quad Algorithm pairs & 22 & 17 & 5 \\
    \quad Souffl\'e benchmark programs & 4 & 3 & 1 \\
    \emph{LLM-generated programs} & 11 & 8 & 3 \\
    \midrule
    Total & 86 & 69 & 17 \\
    \bottomrule
  \end{tabular}
\end{table}

\begin{table}[!htbp]
  \centering
  \captionsetup{skip=4pt}
  \caption{The six MCP tools and six skills available to the proposer in the extended configurations.}
  \label{tab:exp-tools}
  \begin{tabular}{@{}l>{\raggedright\arraybackslash}p{0.64\columnwidth}@{}}
    \toprule
    \multicolumn{2}{@{}l}{\emph{Read-only MCP tools (\WhielTools, \WhielToolsSkills)}} \\
    \midrule
    \toolname{Countermodel} & The finite database behind a refuted check, on which the check's premises hold and its goal fails. \\
    \toolname{Strongest refutations} & For one clause, the refutations whose premise sets are inclusion-maximal. \\
    \toolname{History} & One clause's record: its status, its levels, and every check with its outcome. \\
    \toolname{Ledger} & The chronology of all checks in the run. \\
    \toolname{Validate clauses} & Whether draft clauses parse and fit the schema, without spending a round. \\
    \toolname{Evaluate clauses} & The truth value of draft clauses on retained countermodels or supplied finite instances. \\
    \midrule
    \multicolumn{2}{@{}l}{\emph{Skills: curated tutorials (\WhielSkills, \WhielToolsSkills)}} \\
    \midrule
    \toolname{Ladder strategy} & Build an invariant in layers: level-0 facts, then clauses that need facts about the exit state, then the postcondition over prophecy variables. \\
    \toolname{Bound and frame} & Support a bound on a relation that the loop updates with frame facts about the inputs that the loop never changes. \\
    \toolname{Lagged closure} & State a closure fact that holds only at exit over the previous-iteration snapshot of the relation or over its prophecy variable. \\
    \toolname{Delta coupling} & In a semi-naive loop, relate one step of the accumulator to the accumulator plus the frontier. \\
    \toolname{Mutual simulation} & For two synchronized programs, add one-step simulation containments in both directions. \\
    \toolname{Reading a countermodel} & Turn a countermodel into a clause that excludes it. \\
    \bottomrule
  \end{tabular}
\end{table}

\subsection{\texorpdfstring{Tools and Skills (Section~\ref{subsec:exp-setup})}{Tools and Skills}}\label{app:tools-skills}
Table~\ref{tab:exp-tools} describes the tools and skills summarized in Section~\ref{subsec:exp-setup}.

\subsection{\texorpdfstring{Tasks Solved Only by Stronger Models (Section~\ref{sec:rq3})}{Tasks Solved Only by Stronger Models}}
\label{app:rq3}

\input{figures/rq3-first-model}

Figure~\ref{fig:rq3-first-model} shows which model first solves each of the 17 tasks that GPT-5.5 leaves unsolved in at least one configuration.
For 12 of them, the first model that solves the task depends on the configuration, and \WhielSkills needs a stronger model least often: with skills, GPT-5.5 solves 8 of the 17, against 4 or 5 in the other configurations.
Four tasks resist GPT-5.5 and Sol in every configuration, and Astra solves three of them: \emph{Red/blue alternating paths: linear vs.\ nonlinear} (0134), \emph{DRed deletion vs.\ recomputation} (5016) and \emph{Souffl\'e topological ordering: naive vs.\ semi-naive} (5011).
The semi-naive translation in 5011 is not equivalent to the naive one; GPT-5.5 and Sol mostly propose invariant clauses and run out of time, whereas Astra submits a counterexample in its first model request in all four configurations, and the task is solved after 86--102\,s.
No model solves \emph{Warshall sweep vs.\ PostgreSQL recursive union} (5034), whose invariant nests terminal values two levels deep (Section~\ref{sec:rq4}).

\subsection{\texorpdfstring{Effect of Evaluation Strategy (Section~\ref{sec:rq4})}{Effect of Evaluation Strategy}}
\label{app:rq4}

Section~\ref{sec:rq4} considers one property of a task, whether its reference invariant uses prophecy variables; here we add a second, whether the task verifies an evaluation strategy.

A task \textbf{verifies an evaluation strategy} if it compares evaluation strategies of a recursive query, or checks a non-naive one against the least fixpoint of the query's rules (column E in Table~\ref{tab:app-tasks}); like the use of prophecy variables, this property corresponds to the difficulty of the task.
The success rates of \Whiel with different models on the 69 valid tasks are shown below:

{
    \centering
\small
    \begin{tabular}{|c|c|c|c|c|}
    \hline
        & GPT 5.5 & GPT 5.6 & GPT 6  & Unsolved\\\hline
        Eval.\ strategy (total 25) &  17  (68\%) & 5  (20\%) & 2 (8\%) & 1 (4\%) \\
        Other (total 44) &  40  (91\%) & 3  (7\%) & 1 (2\%) & - \\\hline
    \end{tabular}
}

\noindent
\textbf{Different evaluation strategies.}
GPT-5.5 solves 17 of the 25 tasks that verify an evaluation strategy (68\%) and 40 of the other 44 (91\%); 8 of the 12 valid tasks that it leaves unsolved are among the 25.
In the task \emph{NetworkX per-source TC vs.\ PostgreSQL}, NetworkX completes the closure one source at a time, while PostgreSQL's recursive union extends the newest paths by one edge per iteration; midway, neither result contains the other.
The invariant must track each algorithm's own bookkeeping (NetworkX's frontier, visited set and remaining sources) and connect the two sides through terminal values (e.g., every path that PostgreSQL has found so far lies in the closure that NetworkX eventually returns).
The invariants that GPT-5.5 finds for these 25 tasks are about twice as large as those for the other 44 (median 17 clauses vs.\ 8).

\noindent
\textbf{Invariants over final values.}
GPT-5.5 solves 30 of the 41 tasks whose reference invariant uses prophecy variables (73\%) and 27 of the other 28 (96\%), and 11 of the 12 valid tasks that it leaves unsolved are among the 41.
Different evaluation strategies are a main source of such invariants: 20 of the 25 tasks that verify an evaluation strategy are among the 41.
In the NetworkX task, for example, the clause that connects the two sides is over terminal values.
Of the two tasks whose invariants nest such facts (Section~\ref{sec:rq4}), \emph{Warshall sweep vs.\ PostgreSQL recursive union} (Appendix~\ref{app:benchmark}) needs an invariant stating that the closure Warshall has computed so far is contained in the result that PostgreSQL eventually returns; this holds only because that result is transitively closed, again a fact about a terminal value.

\balance

%% file: figures/rq3-first-model.tex
\begin{figure}[t]
  \centering
  \includegraphics[width=\columnwidth]{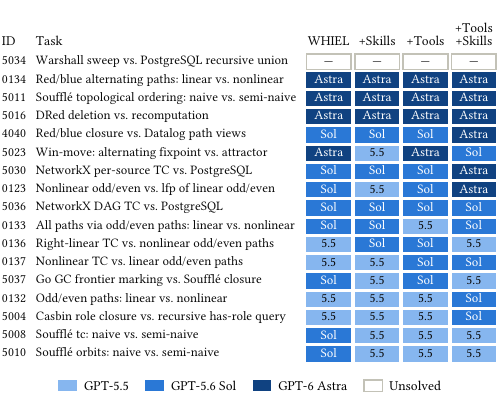}
  \caption{First model that solves each of the 17 tasks that GPT-5.5 leaves unsolved in at least one configuration. No model solves
  task 5034.}
  \label{fig:rq3-first-model}
\end{figure}%

%% file: sections/s8-contributions.tex
\section*{Contribution Statements}

The first three student authors are listed alphabetically by last name. This section delineates their individual contributions to \Whiel\ for the purpose of their dissertation credit.
\begin{itemize}[leftmargin=*]
\itemsep0em
    \item 
{\bf Jesse Comer} designed \Whiel's overall architecture and developed its theoretical and algorithmic foundations. These include the iterative search process characterized by proposer-verifier interaction, the separate certifier pipeline, the use of prophecy variables, the \LH algorithm, and the \AgentHoudini proposer with countermodel feedback. He implemented \Whiel's \Lean library, Rust-based \LH engine and verifier with \Vampire integration, and the \VampLean-based certifier with modifications to ensure kernel-checked SAT proofs. He contributed substantially to the implementation of the \AgentHoudini proposal framework. He also certified the reference benchmark answers and conducted and analyzed the certification experiments.

\item {\bf Fangzhu Shen} designed the agent harness for \Whiel's LLM-based proposer, including countermodel feedback through prompts and MCP tools, the skill library, and the prompts. Fangzhu implemented the proposer from its first version onward, building the initial proposer--verifier loop. Fangzhu also built the experiment runner, contributed to the benchmark, and, with Shuo Zhang, conducted and analyzed the experiments.

\item {\bf Shuo Zhang} led the design and construction of the benchmark. Shuo collected programs from the database literature, open benchmarks and production systems, expressed them in Datalog, and encoded them as verification tasks in Whiel. Shuo also helped implement an early prototype of the Lean--Vampire interface, built and ran the baselines, and, with Fangzhu Shen, conducted and analyzed the experiments.

\end{itemize}
\clearpage